\documentclass[prl,twocolumn,showpacs,preprintnumbers,amsmath,amssymb]{revtex4-1}
\usepackage{graphicx}% Include figure files
\usepackage{dcolumn}% Align table columns on decimal point
\usepackage{bm}% bold math
\begin{document}

\newcommand{\spinup}{\protect{$ \left|\uparrow \right\rangle$}}
\newcommand{\spindown}{\protect{$ \left|\downarrow \right\rangle$}}
\newcommand{\upeq}{\protect{\left | \uparrow \right\rangle}}
\newcommand{\downeq}{\protect{\left | \downarrow \right\rangle}}
\newcommand{\upxeq}{\protect{\left | \uparrow_x \right\rangle}}
\newcommand{\downxeq}{\protect{\left | \downarrow_x \right\rangle}}

\newcommand{\uu}{\protect{$ \left|\uparrow\uparrow\right\rangle$}}
\newcommand{\dd}{\protect{$ \left|\downarrow\downarrow \right\rangle$}}
\newcommand{\ud}{\protect{$ \left|\uparrow\downarrow \right\rangle$}}
\newcommand{\du}{\protect{$ \left|\downarrow\uparrow \right\rangle$}}
\newcommand{\uue}{\protect{\left|\uparrow\uparrow\right\rangle}}
\newcommand{\dde}{\protect{\left|\downarrow\downarrow \right\rangle}}
\newcommand{\ude}{\protect{\left|\uparrow\downarrow \right\rangle}}
\newcommand{\due}{\protect{\left|\downarrow\uparrow \right\rangle}}

\title{3D Sisyphus Cooling of Trapped Ions}

\author{S. Ejtemaee and P.~C. Haljan}
%\email{------}
\affiliation{Department of Physics, Simon Fraser University,
Burnaby, BC, V5A 1S6}

\date{March 1, 2016}% It is always \today, today,

\begin{abstract}
Using a laser polarization gradient, we realize 3D Sisyphus cooling of
$^{171}$Yb$^+$ ions confined in and near the Lamb-Dicke regime in a linear
Paul trap. The cooling rate and final mean motional energy of a single ion
are characterized as a function of laser intensity and compared to
semiclassical and quantum simulations. Sisyphus cooling is also applied to a
linear string of four ions to obtain a mean energy of 1--3 quanta for all
vibrational modes, an approximately order--of--magnitude reduction below
Doppler cooled energies. This is used to enable subsequent, efficient
sideband laser cooling.
\end{abstract}


\maketitle

%\section{\label{sec:intro} Introduction\protect\\}
%\noindent\textbf{Introduction-}

Applications of laser--cooled, trapped ions range from quantum information
processing~\cite{Cirac1995a, Wineland1998a, Blatt2008a, Johanning2009a,
Islam2011a, Blatt2012a} and spectroscopy and metrology~\cite{Schneider2005a,
Roos2006a, Rosenband2008a, Hosaka2009a} to the study of interactions with
cold atoms~\cite{Grier2009a, Zipkes2010a, Haze2015a} and the study of
few-body ``phase transitions"~\cite{Ejtemaee2013a, Pyka2013a, Ulm2013a,
Plenio2013a, Bylinskii2015a}. Central to many of these applications is the
manipulation of the collective vibrational modes of a string of
Coulomb-coupled ions. The modes of interest are often required to be prepared
in their quantum mechanical ground state, which is commonly achieved with
sideband laser cooling~\cite{Diedrich1989a,Monroe1995a, King1998a} or
electromagnetically induced transparency (EIT) cooling~\cite{Roos2000a,
Lin2013}. In practice, these techniques are implemented for reasons of
efficiency in the Lamb-Dicke regime, where the ions' residual amplitude of
vibration is small compared to the wavelength of the cooling
laser~\cite{Wineland1998a, Leibfried2003a, Eschner2003a}. Doppler laser
pre-cooling is usually sufficient to attain this condition, but if the trap
is somewhat weaker, the ions will not begin close to the ground state, or
deep in the Lamb-Dicke regime. In the case of Raman-transition sideband
cooling~\cite{Monroe1995a}, this lengthens and complicates the sequence to
walk the vibrational modes down the ladder of energy levels. Here we consider
Sisyphus laser cooling~\cite{Dalibard1989a, Ungar1989a}, well known for
neutral atoms, to act as a bridge between Doppler and ground-state laser cooling
for ions. This relaxes the requirement on trapping strength, which
is of technological relevance for larger mass ions, and for the weaker axial
confinement necessary to maintain linear strings of larger ion number.
As a convenient means to reach near-ground-state energies, Sisyphus cooling of
trapped ions is a technique of potential broad applicability in
analogy with experiments with neutral atoms.

Since Sisyphus cooling was first demonstrated in a 3D optical
molasses~\cite{Lett1988a}, the technique has been widely adopted to cool
neutral atomic gasses to sub-Doppler temperatures~\cite{Metcalf}. Sisyphus
cooling, primarily due to polarization gradients, has also been used for the
cooling and localization of atoms in optical lattices~\cite{Jessen1992,
Verkerk1992, Kastberg1995, Winoto1999, Bakr2010}, optical
cavities~\cite{Nuszmann2005a, Boozer2006} and optical
tweezers~\cite{schlosser2001, Kaufman2012a, McGovern2011}. Several
theoretical investigations, both semiclassical and quantum, have extended the
concept of Sisyphus cooling to a single ion confined in the Lamb-Dicke
regime, with proposals considering cooling in both
intensity~\cite{Wineland1992a,Cirac1992a} and polarization~\cite{Cirac1993a,
Yoo1993a} gradients. Semiclassical simulations have also been used to study
the final cooling energy in the crossover from the case of a bound ion in the
Lamb-Dicke limit to the free-particle case~\cite{Li1993a}. Despite these
theoretical works, the Sisyphus cooling of trapped ions has been reported only
once, for one and two ions~\cite{Birkl1994a}. In this case, however, the
confinement along the axis being cooled was so weak that the cooling was
essentially the same as for free atoms.

\begin{figure*}
\centering
\includegraphics[width=1.0\linewidth,clip]{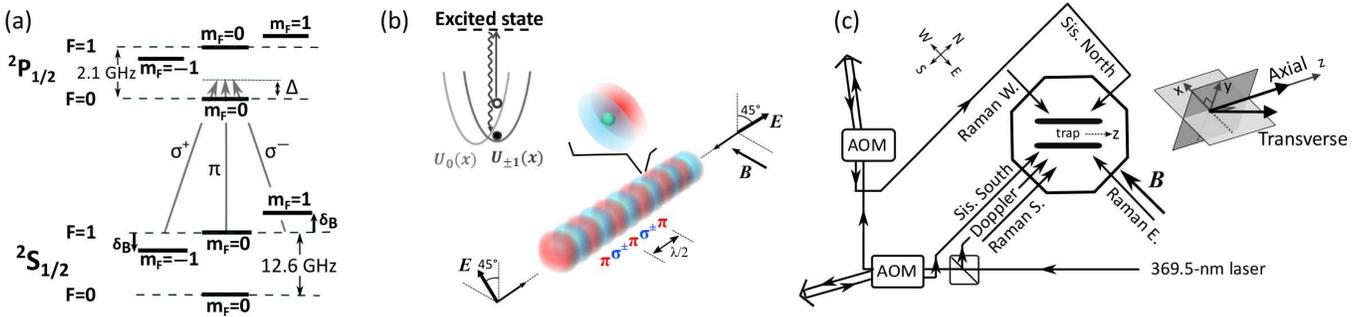}
\caption{(color online) (a) Relevant energy levels of $^{171}$Yb$^+$ for
Sisyphus cooling. (b) An example of a single-ion Sisyphus cooling event from
optical pumping between shifted harmonic trapping potentials associated with
two effective sub--levels of the \protect{$^2$S$_{1/2}|F=1\rangle$} ground
state. The shift in potentials is due to dipole forces from the 1D
polarization gradient shown, which is formed by counter-propagating, linearly
polarized Sisyphus beams in a transverse magnetic field. (c) Laser beam
configuration: Double-pass acousto-optic modulators (AOMs) are used to
control the power and frequency of the Sisyphus beams, which enter the vacuum
chamber and trap from the north and south sides. Perpendicular Raman beams
entering from the east and south are used to probe ion motion in the axial
($\hat{z}$) trap direction; west and south beams are used for the transverse
($\hat{x}$ and $\hat{y}$) directions. Shown at right are the axial and
transverse Raman wavevector directions overlaid on the principal axes of the
trap.} \label{fig:wide}
\end{figure*}

In this Letter we realize the 3D Sisyphus cooling of ions confined in and
near the Lamb-Dicke regime. We first characterize the cooling, based on a
polarization gradient, as a function of laser intensity for a single ion. We
then extend the technique to a linear string of four ions to demonstrate
simultaneous cooling of all its vibrational modes. For our case of
$^{171}$Yb$^+$ ions with an $\protect{F=1\rightarrow F=0}$ cooling transition
[Fig.~\ref{fig:wide}(a)], we construct a periodic polarization gradient in a
transverse magnetic field as shown in
Fig.~\ref{fig:wide}(b)~\cite{Straten1993a}. For high enough magnetic field
(low enough intensity), the ground state coherences associated with coherent
population trapping~\cite{Berkeland2002a, Ejtemaee2010a} can be ignored. A
polarization gradient at the ion trap then gives rise to state-dependent
light shift potentials and spatially dependent optical pumping such that a
Sisyphus cooling effect occurs for blue detuning ($\Delta>0$). A single-ion
cooling limit corresponding to a mean motional quantum number of
$\protect{\bar{n}\approx 1}$ is expected when the depth of the light shift
potentials is on the order of the zero-point energy in the harmonic
trap~\cite{Cirac1993a, Li1993a}.

%%%%%%%%%%%%%%%%%%%%%%%%%%%%%%%%%%%%%%%%%%%%%%%%%%%%%%%%%%%%%%%%%%%%%%%%%%
%fig1: Experiment setup
%%%%%%%%%%%%%%%%%%%%%%%%%%%%%%%%%%%%%%%%%%%%%%%%%%%%%%%%%%%%%%%%%%%%%%%%%%

%\section{\label{sec:experiment} Experiment\protect\\}
%\noindent\textit{Experiment -}

Our detailed studies of Sisyphus cooling with a single $^{171}$Yb$^+$ ion are
done in a linear Paul trap operating at a radio frequency of $\protect{\Omega_T/2\pi=17.4}$
MHz~\cite{Ejtemaee2010a}. Typical secular trap frequencies
are $\protect{\{\omega_x, \omega_y\}/2\pi=\{0.790, 0.766\}}$ MHz in the transverse
direction and $\protect{\omega_z/2\pi=0.525}$ MHz in the axial direction. An
applied 5.9--G magnetic field gives a Zeeman shift of $\protect{\delta_B/2\pi=8.2}$ MHz
between the $\protect{6s^2\!S_{1/2}|F=1,m_F\rangle}$ sub-levels. A laser beam
detuned by --10 MHz of the
$\protect{6s^2\!S_{1/2}(F=1)}-\protect{6p^2\!P_{1/2}(F=0)}$ transition at
$\lambda = 369.5$ nm provides fluorescence detection and Doppler cooling with
$\protect{\bar{n}\sim20}$ in all trap directions. Optical pumping for initialization of
the ion into the $\protect{^2\!S_{1/2}}|0, 0\rangle$ state is achieved with a
laser modulation sideband driving the
$\protect{^2\!S_{1/2}(F=1)}-\protect{^2\!P_{1/2}(F=1)}$ transition.

A polarization--gradient field overlapping the trap is created by two
counter-propagating and cross-polarized laser beams with $\sim$40-$\mu$m
waists. The beams are derived from the Doppler cooling laser
[Fig.~\ref{fig:wide}(c)]. Acousto-optic modulators are used to obtain a
detuning of $\protect{\Delta/2\pi=310}$ MHz above the
$\protect{^2\!S_{1/2}(F=1)}-\protect{^2\!P_{1/2}(F=0)}$ resonance and allow
for independent adjustment of the power ($<$45 $\mu$W) and frequency of the
Sisyphus beams. The projections of either beam's wavevector along the trap
directions ($\protect{\upsilon=\{x,y,z\}}$) have magnitudes
$k_\upsilon=\frac{2\pi}{\lambda}\{\frac{1}{2},\frac{1}{2},\frac{1}{\sqrt{2}}\}$
such that cooling is provided in 3D. The Lamb-Dicke parameters
$\protect{\eta_\upsilon=k_\upsilon r_\upsilon}$ in terms of the ground-state
sizes $r_\upsilon=(\hbar/2m\omega_\upsilon)^{1/2}$ are \{0.052, 0.053,
0.090\}. The polarization of each beam is calibrated \textit{in situ} from ac
Stark shifts measured using microwave Ramsey interferometry between the
$\protect{^2\!S_{1/2}|0,0\rangle}\equiv\downeq$ and
$\protect{^2\!S_{1/2}|1,0\rangle}\equiv\upeq$ states. The single-beam
intensities $I$, or equivalently on-resonant saturation parameters
$s_0=I$/(51 mW/cm$^2$), are also determined in this way, and are balanced to
better than $10\%$. During Sisyphus cooling, the ion can be weakly optically
pumped via the $\protect{^2\!P_{1/2}|F=1\rangle}$ state into the dark
\spindown~state and so out of the cooling cycle. To repump the ion, we use a
pulsed sequence consisting of periods of Sisyphus cooling interleaved with
reset operations composed of a 10-$\mu$s optical pumping pulse followed by a
90-$\mu$s microwave $\pi$-pulse to prepare the ion in the \spinup~state. For
optimal Sisyphus cooling of a single ion we should set the maximum of the
polarization gradient at the center of the trap~\cite{Wineland1992a,
Cirac1993a}, which requires interferometric stability between the Sisyphus
beams. Instead, we introduce a 0.080--MHz frequency difference between the
beams to average over their relative phase -- and its slow drifts -- with a
concomitant decrease in cooling rate and increase in cooling limit.

\begin{figure}[t]
\centering
\includegraphics[width=1.0\linewidth]{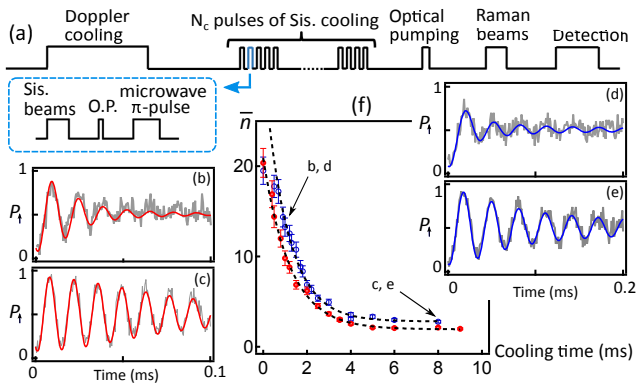}
\caption{(color online) (a) Experiment sequence for assessing Sisyphus
cooling of a single trapped ion. Each step of the $N_c$--pulse Sisyphus sequence
involves a cooling pulse and a reset (i.e. repump) to the \spinup~state using
optical pumping (OP) and microwave pulses. (b--e) Raman carrier Rabi
oscillations in the (b)--(c) transverse  and (d)--(e) axial directions with
cooling times of [(b),(d)] $10\times0.1$ ms and [(c),(e)] $80\times0.1$ ms.
Gray lines are the measured probability $P_\uparrow$ of obtaining \spinup~averaged
over 50 runs per time value. Red and blue lines are fits to extract
$\bar{n}$. (f) Cooling dynamics for transverse (blue) and axial (red)
directions at $s_0=11$. Error bars are statistical uncertainties from fits.
Dashed lines are exponential fits used to extract the
cooling rate. Only points with $\bar{n}\leq15$ are considered to omit the initial
cooling dynamics.
} \label{fig:dynamics}
\end{figure}

The Sisyphus cooling is assessed with thermometry based on motion-sensitive,
two-photon carrier transitions, for example
$\downeq|n_z\rangle\leftrightarrow\upeq|n_z\rangle$~\cite{Wineland1998a}.
A set of three off-resonant Raman beams (detuned by 100 GHz) allows us to
obtain a carrier transition that is sensitive to motion in either the axial
or transverse direction (see Fig.~\ref{fig:wide}(c)). The experiment sequence
[Fig.~\ref{fig:dynamics}(a)] involves 6.6 ms of Doppler cooling, then
Sisyphus cooling, and finally thermometry operations. The thermometry
involves the acquisition of a carrier Rabi oscillation with initialization
via optical pumping to \spindown, and internal-state readout via
state-sensitive fluorescence detection.

%\section{\label{sec:results} Results\protect\\}
%\noindent\textit{Cooling rate -}

We first measure the Sisyphus cooling rate as a function of laser
intensity. The cooling rate at each intensity value is extracted from a set
of measurements of $\bar{n}$ at different Sisyphus cooling times, where the
value of $\bar{n}$ at each time is obtained from a fit to the carrier Rabi
oscillation [Figs.~\ref{fig:dynamics}(b)--\ref{fig:dynamics}(e)]. We vary the Sisyphus cooling time
by varying the number of Sisyphus pulses with their duration kept constant. The
pulse duration for a given beam intensity is set to keep the probability of
pumping out of the cooling cycle to 15\%. The fit function for the Rabi
oscillation assumes an initial thermal distribution of motional Fock states
and includes fixed corrections for detection efficiencies. The only free fit parameters are $\bar{n}$ and a
carrier Rabi frequency scale. In the transverse direction, our Raman setup
couples to both $x$ and $y$ motions with equal Raman wavevector
projection onto each axis [Fig.~\ref{fig:wide}(c)]. We use an approximate 2D model for the transverse
fits in which we assume the same $\bar{n}$ for both axes and ignore the
effect of Raman transitions related to cross-mode coupling between the
axes~\cite{Wineland1998a}. Even though we do not resolve the closest of these
transitions to the carrier, simulations show that the fit model is adequate
for the $\bar{n}$ range considered ($\leq$5\% systematic effect at highest
$\bar{n}$ values)~\cite{Ejtemaee2016b}.

\begin{figure}[t]
\centering
\includegraphics[width=1.0\linewidth]{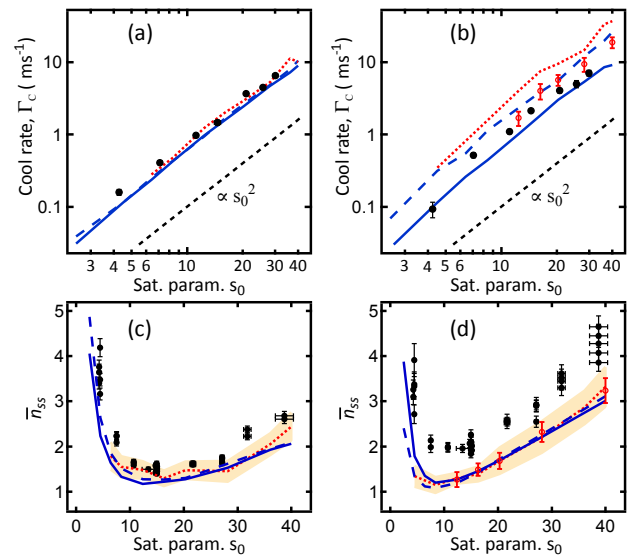}
\caption{(color online) Sisyphus cooling rate $\Gamma_c$ (top) and
steady-state mean phonon number $\bar{n}_{ss}$ (bottom) as a function of
single-beam saturation parameter $s_0$ for the transverse direction (left)
and axial direction (right). Panels include experimental data (black filled
circles) and predictions from 1D (dashed blue line) and 3D (solid blue line)
semiclassical simulations, and 1D quantum simulations with initial
\protect{$n_i=8$} Fock state (dotted red line) and initial $\bar{n} = 22$
thermal state (red circles). Semiclassical simulations average over 1000
Monte Carlo runs. Quantum simulations average over 40 (80) runs for Fock
(thermal) initial states. Vertical error bars for data are statistical
uncertainties from fits, and horizontal error bars account for calibration
uncertainty and drifts in laser intensity. In (b), error bars for the quantum
simulation with thermal initial state are bootstrap uncertainties. Shaded error
bands and error bars for quantum simulations in (c) and (d) show the standard
deviation of fluctuations at steady state.} \label{fig:nbar-coolrate}
\end{figure}

Typical cooling dynamics at $\protect{s_0=11}$ are shown in
Fig.~\ref{fig:dynamics}(f) for both the axial and transverse directions. An
exponential fit is used to extract a cooling time constant $\tau$. Since on
average the ion is cooled $85\%$ of the time, due to the effect of pumping
dark, our plotted cooling rate is calculated as $\Gamma_c=(\tau/0.85)^{-1}$.
Figures~\ref{fig:nbar-coolrate}(a)--\ref{fig:nbar-coolrate}(b) show the
intensity dependence of the axial and transverse cooling rates. The cooling
rates in the two directions compare within a factor of 2 of each other over
the measured range, spanning more than a factor of 40. Power-law fits of the
axial and transverse data give exponents 1.98(6) and 1.91(3) respectively,
which match well with the expected $s_0^2$ scaling in the Lamb-Dicke regime
and in the absence of coherences between Zeeman levels~\cite{Cirac1993a,
Ejtemaee2016b}.

%%%%%%%%%%%%%%%%%%%%%%%%%%%%%%%%%%%%%%%%%%%%%%%%%%%%%%%%%%%%%%%%%%%%%%%%%%
%fig2: Single ion carrier tscan and cooling rate decay plot
%%%%%%%%%%%%%%%%%%%%%%%%%%%%%%%%%%%%%%%%%%%%%%%%%%%%%%%%%%%%%%%%%%%%%%%%%%

%%%%%%%%%%%%%%%%%%%%%%%%%%%%%%%%%%%%%%%%%%%%%%%%%%%%%%%%%%%%%%%%%%%%%%%%%%
%fig3: Cooling rate and final temperature vs beam intensity
%%%%%%%%%%%%%%%%%%%%%%%%%%%%%%%%%%%%%%%%%%%%%%%%%%%%%%%%%%%%%%%%%%%%%%%%%%

%\noindent\textit{Steady-state energy -}

Next, we measure the steady-state mean occupation number $\bar{n}_{ss}$ as a
function of laser intensity. Potential heating sources such as micro-motion
and laser power noise are checked and minimized (if necessary) on a regular
basis during data collection. At each intensity, the single-pulse cooling
time is set to keep the probability of pumping out of the cooling cycle at a
fixed value of 20\%, and a cooling time in excess of 9$\tau$ is chosen to
allow the energy of the ion to reach equilibrium.
Figures~\ref{fig:nbar-coolrate}(c)--\ref{fig:nbar-coolrate}(d) show the
intensity dependence of $\bar{n}_{ss}$ for both the axial and transverse
directions. In each case, a cooling limit of $\bar{n}_{ss}\simeq$ 1.5--2 is
obtained at an optimum intensity. The lower optimum intensity in the less
strongly confined axial direction is consistent with the theoretical
expectation for an ion in the Lamb-Dicke regime~\cite{Wineland1992a,
Cirac1993a}.

%\noindent\textit{Comparison to theory -}

Both semiclassical and quantum simulations are performed to assess the
experimental results. For all simulations the trap is treated in the
pseudo-potential approximation, and a 0.080--MHz frequency difference between
the Sisyphus beams is included. We consider both 1D and 3D semiclassical
Monte Carlo simulations, which treat the motion of the ion classically and
include a period of Doppler cooling followed by Sisyphus cooling to match the
experiment. The Sisyphus cooling model follows the rate-equation approach
of~\cite{Li1993a} with the appropriate diffusion heating terms calculated
according to~\cite{Gordon1980a,Nienhuis1991a}. The effect of photon
scattering from the $^2P_{1/2}|F=1\rangle$ states, which is omitted in the
simulations, is included in the $\bar{n}_{ss}$ values presented through an
intensity--dependent correction determined analytically~\cite{Ejtemaee2016b}.
The quantum simulation is implemented in 1D with the Monte-Carlo wavefunction
method~\cite{Dalibard1992a} according to~\cite{Johansson20131234}. It
includes the hyperfine structure of the $^2S_{1/2}$--$^2P_{1/2}$ transition
and coherences between Zeeman levels, but ignores any coherences between
$F$--levels. The \protect{$^2S_{1/2}|0,0\rangle$} state is effectively
eliminated by assuming an instantaneous recoilless repump. The majority of
our quantum simulations use an initial Fock state of $n_i=8$ and are limited
to a Hilbert space of 20 motional $n$--levels in order to restrict the
computational time required. A limited subset of points is repeated with a
thermal initial state at Doppler temperature and a Hilbert space of 200
$n$--levels.

For the transverse cooling rate~[Fig.~\ref{fig:nbar-coolrate}(a)], all the
simulation models match the experimental results fairly well. In the weaker
axial direction~[Fig.~\ref{fig:nbar-coolrate}(b)], the 3D semiclassical
simulation matches the data better overall than the 1D semiclassical
simulation and distinctly better than the quantum simulation with $n_i=8$.
The discrepancy between the 1D and 3D semiclassical simulations (by a factor
of 2--3) suggests that the axial cooling behavior is affected by the
transverse motion, perhaps due to motional coupling or due to the additional
delocalization of the ion. Simulations in a tighter trap by a factor of three
(that is deeper in the Lamb-Dicke regime) do not show this difference. The
higher axial cooling rate predicted by the 1D quantum simulation with
$\protect{n_i=8}$, by an overall factor of 3--4, is related to the lower
initial motional energy used in the calculation. As shown in
Fig.~\ref{fig:nbar-coolrate}(b), a thermal initial state with a Doppler
cooled value of $\protect{\bar{n}=22}$ brings the quantum result in line with
the 1D semiclassical simulation and closer to the experimental data,
indicating the effect of deviations from the Lamb-Dicke regime in the early
cooling dynamics.

For the transverse $\bar{n}_{ss}$ in
Fig.~\ref{fig:nbar-coolrate}(c), the quantum and semiclassical simulations
lie close to one another and only show a small discrepancy with the data over the
intensity range considered. In the axial direction, there is a much stronger
discrepancy between the experiment and theory by up to a factor of 2,
although the general behaviors still agree. While the source of the discrepancies remains
to be identified, we have verified that the carrier thermometry does
not present a measurement limit.

%%%%%%%%%%%%%%%%%%%%%%%%%%%%%%%%%%%%%%%%%%%%%%%%%%%%%%%%%%%%%%%%%%%%%%%%%%
%fig4: Sisyphus and sideband  cooling of 4 ions
%%%%%%%%%%%%%%%%%%%%%%%%%%%%%%%%%%%%%%%%%%%%%%%%%%%%%%%%%%%%%%%%%%%%%%%%%%
\begin{figure}[t]
\centering
\includegraphics[width=1.0\linewidth]{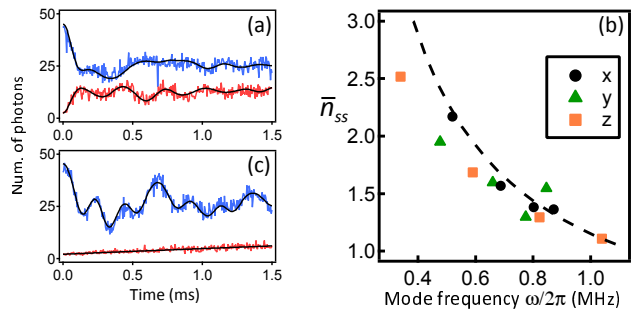}
\caption{(color online) (a) Raman Rabi oscillations on the first red sideband
transition for the $y$--zigzag vibrational mode (\protect{$\omega/2\pi$=0.48}
MHz) in a Sisyphus-cooled linear string of four $^{171}$Yb$^+$ ions. Initial
internal state is either $|\downarrow, \downarrow,
\downarrow,\downarrow\rangle$ (red) or $|\uparrow, \uparrow,
\uparrow,\uparrow\rangle$ (blue). Vertical scale proportional to the number
of ions in \spinup~averaged over 50 runs. Black lines are a combined fit to
extract \protect{$\bar{n}_{ss}=1.95(4)$} and include approximate models for
optical pumping and contrast loss for red and blue curves, respectively. (b)
Mean phonon number for all four-ion vibrational modes following Sisyphus
cooling at $s_0=15$. Trap frequencies \{0.84, 0.87, 0.34\} MHz. Dashed line
shows $\omega^{-1}$ scaling for reference. (c) Same mode as (a) but Sisyphus
then sideband cooled ($\bar{n}\leq0.05$ from fit). } \label{fig:four_ion}
\end{figure}

%\noindent\textit{Four-ion Sisyphus cooling -}

In the final experiment, we extend Sisyphus cooling to a linear string of
ions, specifically $\protect{N=4}$ ions confined in a slightly weaker axial
trap with $\protect{\omega_z/2\pi=0.34}$ MHz. All the ions in the 16-$\mu$m
long string interact with the polarizaton gradient field. We choose
$\protect{s_0=15}$ and apply the cooling for a duration of $30\times0.2$ ms.
The experimental sequence is the same as for a single ion; however, for
thermometry of each vibrational normal mode, we measure red-sideband Rabi
oscillations starting from either the
$\protect{|\downarrow\downarrow\downarrow\downarrow\rangle}$ or the
$\protect{|\uparrow\uparrow\uparrow\uparrow\rangle}$ state, and fit the
oscillations together to find $\bar{n}_{ss}$ for an assumed thermal
distribution~\cite{sidebandthermometry}. The Raman beams nominally uniformly
illuminate the ions. Figure~\ref{fig:four_ion}(a) shows an example of the
Rabi oscillations for the $y$-axis zigzag mode. The fit function ignores
spectator--mode effects~\cite{Wineland1998a}, which are expected to be small
given that all modes are Sisyphus cooled. We also modify the fit function to
account for experimental imperfections in the Raman transition, including
optical pumping as a result of spontaneous emission and loss of contrast due
to residual intensity inhomogeneities across the ion string.
Figure~\ref{fig:four_ion}(b) shows the mean vibrational number $\bar{n}_{ss}$
following Sisyphus cooling for all of the $3N$ normal modes as a
function of their frequency. The Sisyphus cooling reduces the energy of all
modes to $\bar{n}_{ss}\leq3$. Starting from the Sisyphus cooled string, we
have implemented separate sideband cooling of all modes with a typical result
of $\bar{n}\leq0.05$ (for example, Fig.~\ref{fig:four_ion}(c)).

%\noindent\textit{Discussion and conclusions -}

In conclusion, Sisyphus laser cooling has been used to reduce the thermal energy of
trapped ions in 3D by approximately an order of magnitude, thereby bridging
Doppler and sideband cooling in our setup. In addition to providing near
ground-state cooling, Sisyphus cooling benefits from a simplicity and robustness
because it is not a resonant process, and so does not require fine tuning of
multiple cooling parameters. The Sisyphus technique is convenient to
implement since it requires only modest optical power and uses the same
single direction of optical access as for Doppler cooling. In future, the
Sisyphus cooling time could be reduced in our setup by eliminating the pulsed
reset in favor of a continuously active repump laser. Further investigation
of the cooling performance with respect to a range of parameters, and the
cross-over to behavior outside the Lamb-Dicke regime~\cite{Li1993a}, will be
presented elsewhere~\cite{Ejtemaee2016b}. The Sisyphus technique should
immediately extend to ion strings at least moderately larger than four ions.
This opens up the possibility in our setup to explore dynamics of the
linear-zigzag transition in and near the quantum regime~\cite{Retzker2008a},
and should be of interest for recent proposals to study heat transport in ion
strings~\cite{Lin2011, Pruttivarasin2011, Bermudez2013, Freitas2013,
Ruiz2014}. Sisyphus cooling may also be useful in quantum information
applications where ground-state cooling is not required, for example in
microwave-based quantum logic~\cite{Ospelkaus2011, Timoney2011, Webster2013a} or other proposed schemes~\cite{Lin2009a}.

%%\begin{acknowledgments}
%%This work was supported by NSERC and CFI LOF.
%%\end{acknowledgments}

%\bibliographystyle{apsrevmod}
%\bibliography{Sisbib}% Produces the bibliography via BibTeX.

\begin{thebibliography}{64}%
\makeatletter
\providecommand \@ifxundefined [1]{%
 \@ifx{#1\undefined}
}%
\providecommand \@ifnum [1]{%
 \ifnum #1\expandafter \@firstoftwo
 \else \expandafter \@secondoftwo
 \fi
}%
\providecommand \@ifx [1]{%
 \ifx #1\expandafter \@firstoftwo
 \else \expandafter \@secondoftwo
 \fi
}%
\providecommand \natexlab [1]{#1}%
\providecommand \enquote  [1]{``#1''}%
\providecommand \bibnamefont  [1]{#1}%
\providecommand \bibfnamefont [1]{#1}%
\providecommand \citenamefont [1]{#1}%
\providecommand \href@noop [0]{\@secondoftwo}%
\providecommand \href [0]{\begingroup \@sanitize@url \@href}%
\providecommand \@href[1]{\@@startlink{#1}\@@href}%
\providecommand \@@href[1]{\endgroup#1\@@endlink}%
\providecommand \@sanitize@url [0]{\catcode `\\12\catcode `\$12\catcode
  `\&12\catcode `\#12\catcode `\^12\catcode `\_12\catcode `\%12\relax}%
\providecommand \@@startlink[1]{}%
\providecommand \@@endlink[0]{}%
\providecommand \url  [0]{\begingroup\@sanitize@url \@url }%
\providecommand \@url [1]{\endgroup\@href {#1}{\urlprefix }}%
\providecommand \urlprefix  [0]{URL }%
\providecommand \Eprint [0]{\href }%
\providecommand \doibase [0]{http://dx.doi.org/}%
\providecommand \selectlanguage [0]{\@gobble}%
\providecommand \bibinfo  [0]{\@secondoftwo}%
\providecommand \bibfield  [0]{\@secondoftwo}%
\providecommand \translation [1]{[#1]}%
\providecommand \BibitemOpen [0]{}%
\providecommand \bibitemStop [0]{}%
\providecommand \bibitemNoStop [0]{.\EOS\space}%
\providecommand \EOS [0]{\spacefactor3000\relax}%
\providecommand \BibitemShut  [1]{\csname bibitem#1\endcsname}%
\let\auto@bib@innerbib\@empty
%</preamble>
\bibitem [{\citenamefont {Cirac}\ and\ \citenamefont
  {Zoller}(1995)}]{Cirac1995a}%
  \BibitemOpen
  \bibfield  {author} {\bibinfo {author} {\bibfnamefont {J.~I.}\ \bibnamefont
  {Cirac}}\ and\ \bibinfo {author} {\bibfnamefont {P.}~\bibnamefont {Zoller}},\
  }\href {\doibase 10.1103/PhysRevLett.74.4091} {\bibfield  {journal} {\bibinfo
   {journal} {Phys. Rev. Lett.}\ }\textbf {\bibinfo {volume} {74}},\ \bibinfo
  {pages} {4091} (\bibinfo {year} {1995})}\BibitemShut {NoStop}%
\bibitem [{\citenamefont {Wineland}\ \emph {et~al.}(1998)\citenamefont
  {Wineland}, \citenamefont {Monroe}, \citenamefont {Itano}, \citenamefont
  {Leibfried}, \citenamefont {King},\ and\ \citenamefont
  {Meekhof}}]{Wineland1998a}%
  \BibitemOpen
  \bibfield  {author} {\bibinfo {author} {\bibfnamefont {D.}~\bibnamefont
  {Wineland}}, \bibinfo {author} {\bibfnamefont {C.}~\bibnamefont {Monroe}},
  \bibinfo {author} {\bibfnamefont {W.~M.}\ \bibnamefont {Itano}}, \bibinfo
  {author} {\bibfnamefont {D.}~\bibnamefont {Leibfried}}, \bibinfo {author}
  {\bibfnamefont {B.~E.}\ \bibnamefont {King}}, \ and\ \bibinfo {author}
  {\bibfnamefont {D.~M.}\ \bibnamefont {Meekhof}},\ }\href@noop {} {\bibfield
  {journal} {\bibinfo  {journal} {J. Res. Natl. Inst. Stand. and Technol.}\
  }\textbf {\bibinfo {volume} {103}},\ \bibinfo {pages} {259} (\bibinfo {year}
  {1998})}\BibitemShut {NoStop}%
\bibitem [{\citenamefont {Blatt}\ and\ \citenamefont
  {Wineland}(2008)}]{Blatt2008a}%
  \BibitemOpen
  \bibfield  {author} {\bibinfo {author} {\bibfnamefont {R.}~\bibnamefont
  {Blatt}}\ and\ \bibinfo {author} {\bibfnamefont {D.}~\bibnamefont
  {Wineland}},\ }\href {\doibase 10.1038/nature07125} {\bibfield  {journal}
  {\bibinfo  {journal} {Nature}\ }\textbf {\bibinfo {volume} {453}},\ \bibinfo
  {pages} {1008} (\bibinfo {year} {2008})}\BibitemShut {NoStop}%
\bibitem [{\citenamefont {Johanning}\ \emph {et~al.}(2009)\citenamefont
  {Johanning}, \citenamefont {Braun}, \citenamefont {Timoney}, \citenamefont
  {Elman}, \citenamefont {Neuhauser},\ and\ \citenamefont
  {Wunderlich}}]{Johanning2009a}%
  \BibitemOpen
  \bibfield  {author} {\bibinfo {author} {\bibfnamefont {M.}~\bibnamefont
  {Johanning}}, \bibinfo {author} {\bibfnamefont {A.}~\bibnamefont {Braun}},
  \bibinfo {author} {\bibfnamefont {N.}~\bibnamefont {Timoney}}, \bibinfo
  {author} {\bibfnamefont {V.}~\bibnamefont {Elman}}, \bibinfo {author}
  {\bibfnamefont {W.}~\bibnamefont {Neuhauser}}, \ and\ \bibinfo {author}
  {\bibfnamefont {C.}~\bibnamefont {Wunderlich}},\ }\href {\doibase
  10.1103/PhysRevLett.102.073004} {\bibfield  {journal} {\bibinfo  {journal}
  {Phys. Rev. Lett.}\ }\textbf {\bibinfo {volume} {102}},\ \bibinfo {pages}
  {073004} (\bibinfo {year} {2009})}\BibitemShut {NoStop}%
\bibitem [{\citenamefont {Islam}\ \emph {et~al.}(2011)\citenamefont {Islam},
  \citenamefont {Edwards}, \citenamefont {Kim}, \citenamefont {Korenblit},
  \citenamefont {Noh}, \citenamefont {Carmichael}, \citenamefont {Lin},
  \citenamefont {Duan}, \citenamefont {Joseph~Wang}, \citenamefont
  {Freericks},\ and\ \citenamefont {Monroe}}]{Islam2011a}%
  \BibitemOpen
  \bibfield  {author} {\bibinfo {author} {\bibfnamefont {R.}~\bibnamefont
  {Islam}}, \bibinfo {author} {\bibfnamefont {E.}~\bibnamefont {Edwards}},
  \bibinfo {author} {\bibfnamefont {K.}~\bibnamefont {Kim}}, \bibinfo {author}
  {\bibfnamefont {S.}~\bibnamefont {Korenblit}}, \bibinfo {author}
  {\bibfnamefont {C.}~\bibnamefont {Noh}}, \bibinfo {author} {\bibfnamefont
  {H.}~\bibnamefont {Carmichael}}, \bibinfo {author} {\bibfnamefont {G.-D.}\
  \bibnamefont {Lin}}, \bibinfo {author} {\bibfnamefont {L.-M.}\ \bibnamefont
  {Duan}}, \bibinfo {author} {\bibfnamefont {C.-C.}\ \bibnamefont
  {Joseph~Wang}}, \bibinfo {author} {\bibfnamefont {J.}~\bibnamefont
  {Freericks}}, \ and\ \bibinfo {author} {\bibfnamefont {C.}~\bibnamefont
  {Monroe}},\ }\href {\doibase 10.1038/ncomms1374} {\bibfield  {journal}
  {\bibinfo  {journal} {Nat Commun}\ }\textbf {\bibinfo {volume} {2}},\
  \bibinfo {pages} {377} (\bibinfo {year} {2011})}\BibitemShut {NoStop}%
\bibitem [{\citenamefont {Blatt}\ and\ \citenamefont
  {Roos}(2012)}]{Blatt2012a}%
  \BibitemOpen
  \bibfield  {author} {\bibinfo {author} {\bibfnamefont {R.}~\bibnamefont
  {Blatt}}\ and\ \bibinfo {author} {\bibfnamefont {C.~F.}\ \bibnamefont
  {Roos}},\ }\href {\doibase 10.1038/nphys2252} {\bibfield  {journal} {\bibinfo
   {journal} {Nat Phys}\ }\textbf {\bibinfo {volume} {8}},\ \bibinfo {pages}
  {277} (\bibinfo {year} {2012})}\BibitemShut {NoStop}%
\bibitem [{\citenamefont {Schneider}\ \emph {et~al.}(2005)\citenamefont
  {Schneider}, \citenamefont {Peik},\ and\ \citenamefont
  {Tamm}}]{Schneider2005a}%
  \BibitemOpen
  \bibfield  {author} {\bibinfo {author} {\bibfnamefont {T.}~\bibnamefont
  {Schneider}}, \bibinfo {author} {\bibfnamefont {E.}~\bibnamefont {Peik}}, \
  and\ \bibinfo {author} {\bibfnamefont {C.}~\bibnamefont {Tamm}},\ }\href
  {\doibase 10.1103/PhysRevLett.94.230801} {\bibfield  {journal} {\bibinfo
  {journal} {Phys. Rev. Lett.}\ }\textbf {\bibinfo {volume} {94}},\ \bibinfo
  {pages} {230801} (\bibinfo {year} {2005})}\BibitemShut {NoStop}%
\bibitem [{\citenamefont {Roos}\ \emph {et~al.}(2006)\citenamefont {Roos},
  \citenamefont {Chwalla}, \citenamefont {Kim}, \citenamefont {Riebe},\ and\
  \citenamefont {Blatt}}]{Roos2006a}%
  \BibitemOpen
  \bibfield  {author} {\bibinfo {author} {\bibfnamefont {C.~F.}\ \bibnamefont
  {Roos}}, \bibinfo {author} {\bibfnamefont {M.}~\bibnamefont {Chwalla}},
  \bibinfo {author} {\bibfnamefont {K.}~\bibnamefont {Kim}}, \bibinfo {author}
  {\bibfnamefont {M.}~\bibnamefont {Riebe}}, \ and\ \bibinfo {author}
  {\bibfnamefont {R.}~\bibnamefont {Blatt}},\ }\href {\doibase
  10.1038/nature05101} {\bibfield  {journal} {\bibinfo  {journal} {Nature}\
  }\textbf {\bibinfo {volume} {443}},\ \bibinfo {pages} {316} (\bibinfo {year}
  {2006})}\BibitemShut {NoStop}%
\bibitem [{\citenamefont {Rosenband}\ \emph {et~al.}(2008)\citenamefont
  {Rosenband}, \citenamefont {Hume}, \citenamefont {Schmidt}, \citenamefont
  {Chou}, \citenamefont {Brusch}, \citenamefont {Lorini}, \citenamefont
  {Oskay}, \citenamefont {Drullinger}, \citenamefont {Fortier}, \citenamefont
  {Stalnaker}, \citenamefont {Diddams}, \citenamefont {Swann}, \citenamefont
  {Newbury}, \citenamefont {Itano}, \citenamefont {Wineland},\ and\
  \citenamefont {Bergquist}}]{Rosenband2008a}%
  \BibitemOpen
  \bibfield  {author} {\bibinfo {author} {\bibfnamefont {T.}~\bibnamefont
  {Rosenband}}, \bibinfo {author} {\bibfnamefont {D.~B.}\ \bibnamefont {Hume}},
  \bibinfo {author} {\bibfnamefont {P.~O.}\ \bibnamefont {Schmidt}}, \bibinfo
  {author} {\bibfnamefont {C.~W.}\ \bibnamefont {Chou}}, \bibinfo {author}
  {\bibfnamefont {A.}~\bibnamefont {Brusch}}, \bibinfo {author} {\bibfnamefont
  {L.}~\bibnamefont {Lorini}}, \bibinfo {author} {\bibfnamefont {W.~H.}\
  \bibnamefont {Oskay}}, \bibinfo {author} {\bibfnamefont {R.~E.}\ \bibnamefont
  {Drullinger}}, \bibinfo {author} {\bibfnamefont {T.~M.}\ \bibnamefont
  {Fortier}}, \bibinfo {author} {\bibfnamefont {J.~E.}\ \bibnamefont
  {Stalnaker}}, \bibinfo {author} {\bibfnamefont {S.~A.}\ \bibnamefont
  {Diddams}}, \bibinfo {author} {\bibfnamefont {W.~C.}\ \bibnamefont {Swann}},
  \bibinfo {author} {\bibfnamefont {N.~R.}\ \bibnamefont {Newbury}}, \bibinfo
  {author} {\bibfnamefont {W.~M.}\ \bibnamefont {Itano}}, \bibinfo {author}
  {\bibfnamefont {D.~J.}\ \bibnamefont {Wineland}}, \ and\ \bibinfo {author}
  {\bibfnamefont {J.~C.}\ \bibnamefont {Bergquist}},\ }\href {\doibase
  10.1126/science.1154622} {\bibfield  {journal} {\bibinfo  {journal}
  {Science}\ }\textbf {\bibinfo {volume} {319}},\ \bibinfo {pages} {1808}
  (\bibinfo {year} {2008})}\BibitemShut {NoStop}%
\bibitem [{\citenamefont {Hosaka}\ \emph {et~al.}(2009)\citenamefont {Hosaka},
  \citenamefont {Webster}, \citenamefont {Stannard}, \citenamefont {Walton},
  \citenamefont {Margolis},\ and\ \citenamefont {Gill}}]{Hosaka2009a}%
  \BibitemOpen
  \bibfield  {author} {\bibinfo {author} {\bibfnamefont {K.}~\bibnamefont
  {Hosaka}}, \bibinfo {author} {\bibfnamefont {S.~A.}\ \bibnamefont {Webster}},
  \bibinfo {author} {\bibfnamefont {A.}~\bibnamefont {Stannard}}, \bibinfo
  {author} {\bibfnamefont {B.~R.}\ \bibnamefont {Walton}}, \bibinfo {author}
  {\bibfnamefont {H.~S.}\ \bibnamefont {Margolis}}, \ and\ \bibinfo {author}
  {\bibfnamefont {P.}~\bibnamefont {Gill}},\ }\href {\doibase
  10.1103/PhysRevA.79.033403} {\bibfield  {journal} {\bibinfo  {journal} {Phys.
  Rev. A}\ }\textbf {\bibinfo {volume} {79}},\ \bibinfo {pages} {033403}
  (\bibinfo {year} {2009})}\BibitemShut {NoStop}%
\bibitem [{\citenamefont {Grier}\ \emph {et~al.}(2009)\citenamefont {Grier},
  \citenamefont {Cetina}, \citenamefont {Oru\ifmmode \check{c}\else
  \v{c}\fi{}evi\ifmmode~\acute{c}\else \'{c}\fi{}},\ and\ \citenamefont
  {Vuleti\ifmmode~\acute{c}\else \'{c}\fi{}}}]{Grier2009a}%
  \BibitemOpen
  \bibfield  {author} {\bibinfo {author} {\bibfnamefont {A.~T.}\ \bibnamefont
  {Grier}}, \bibinfo {author} {\bibfnamefont {M.}~\bibnamefont {Cetina}},
  \bibinfo {author} {\bibfnamefont {F.}~\bibnamefont {Oru\ifmmode
  \check{c}\else \v{c}\fi{}evi\ifmmode~\acute{c}\else \'{c}\fi{}}}, \ and\
  \bibinfo {author} {\bibfnamefont {V.}~\bibnamefont
  {Vuleti\ifmmode~\acute{c}\else \'{c}\fi{}}},\ }\href {\doibase
  10.1103/PhysRevLett.102.223201} {\bibfield  {journal} {\bibinfo  {journal}
  {Phys. Rev. Lett.}\ }\textbf {\bibinfo {volume} {102}},\ \bibinfo {pages}
  {223201} (\bibinfo {year} {2009})}\BibitemShut {NoStop}%
\bibitem [{\citenamefont {Zipkes}\ \emph {et~al.}(2010)\citenamefont {Zipkes},
  \citenamefont {Palzer}, \citenamefont {Sias},\ and\ \citenamefont
  {Kohl}}]{Zipkes2010a}%
  \BibitemOpen
  \bibfield  {author} {\bibinfo {author} {\bibfnamefont {C.}~\bibnamefont
  {Zipkes}}, \bibinfo {author} {\bibfnamefont {S.}~\bibnamefont {Palzer}},
  \bibinfo {author} {\bibfnamefont {C.}~\bibnamefont {Sias}}, \ and\ \bibinfo
  {author} {\bibfnamefont {M.}~\bibnamefont {Kohl}},\ }\href
  {http://dx.doi.org/10.1038/nature08865} {\bibfield  {journal} {\bibinfo
  {journal} {Nature}\ }\textbf {\bibinfo {volume} {464}},\ \bibinfo {pages}
  {388} (\bibinfo {year} {2010})}\BibitemShut {NoStop}%
\bibitem [{\citenamefont {Haze}\ \emph {et~al.}(2015)\citenamefont {Haze},
  \citenamefont {Saito}, \citenamefont {Fujinaga},\ and\ \citenamefont
  {Mukaiyama}}]{Haze2015a}%
  \BibitemOpen
  \bibfield  {author} {\bibinfo {author} {\bibfnamefont {S.}~\bibnamefont
  {Haze}}, \bibinfo {author} {\bibfnamefont {R.}~\bibnamefont {Saito}},
  \bibinfo {author} {\bibfnamefont {M.}~\bibnamefont {Fujinaga}}, \ and\
  \bibinfo {author} {\bibfnamefont {T.}~\bibnamefont {Mukaiyama}},\ }\href
  {\doibase 10.1103/PhysRevA.91.032709} {\bibfield  {journal} {\bibinfo
  {journal} {Phys. Rev. A}\ }\textbf {\bibinfo {volume} {91}},\ \bibinfo
  {pages} {032709} (\bibinfo {year} {2015})}\BibitemShut {NoStop}%
\bibitem [{\citenamefont {Ejtemaee}\ and\ \citenamefont
  {Haljan}(2013)}]{Ejtemaee2013a}%
  \BibitemOpen
  \bibfield  {author} {\bibinfo {author} {\bibfnamefont {S.}~\bibnamefont
  {Ejtemaee}}\ and\ \bibinfo {author} {\bibfnamefont {P.~C.}\ \bibnamefont
  {Haljan}},\ }\href {\doibase 10.1103/PhysRevA.87.051401} {\bibfield
  {journal} {\bibinfo  {journal} {Phys. Rev. A}\ }\textbf {\bibinfo {volume}
  {87}},\ \bibinfo {pages} {051401} (\bibinfo {year} {2013})}\BibitemShut
  {NoStop}%
\bibitem [{\citenamefont {Pyka}\ \emph {et~al.}(2013)\citenamefont {Pyka},
  \citenamefont {Keller}, \citenamefont {Partner}, \citenamefont {Nigmatullin},
  \citenamefont {Burgermeister}, \citenamefont {Meier}, \citenamefont
  {Kuhlmann}, \citenamefont {Retzker}, \citenamefont {Plenio}, \citenamefont
  {Zurek}, \citenamefont {del Campo},\ and\ \citenamefont
  {Mehlst\"aubler}}]{Pyka2013a}%
  \BibitemOpen
  \bibfield  {author} {\bibinfo {author} {\bibfnamefont {K.}~\bibnamefont
  {Pyka}}, \bibinfo {author} {\bibfnamefont {J.}~\bibnamefont {Keller}},
  \bibinfo {author} {\bibfnamefont {H.~L.}\ \bibnamefont {Partner}}, \bibinfo
  {author} {\bibfnamefont {R.}~\bibnamefont {Nigmatullin}}, \bibinfo {author}
  {\bibfnamefont {T.}~\bibnamefont {Burgermeister}}, \bibinfo {author}
  {\bibfnamefont {D.~M.}\ \bibnamefont {Meier}}, \bibinfo {author}
  {\bibfnamefont {K.}~\bibnamefont {Kuhlmann}}, \bibinfo {author}
  {\bibfnamefont {A.}~\bibnamefont {Retzker}}, \bibinfo {author} {\bibfnamefont
  {M.~B.}\ \bibnamefont {Plenio}}, \bibinfo {author} {\bibfnamefont {W.~H.}\
  \bibnamefont {Zurek}}, \bibinfo {author} {\bibfnamefont {A.}~\bibnamefont
  {del Campo}}, \ and\ \bibinfo {author} {\bibfnamefont {T.~E.}\ \bibnamefont
  {Mehlst\"aubler}},\ }\href {http://dx.doi.org/10.1038/ncomms3291} {\bibfield
  {journal} {\bibinfo  {journal} {Nat Commun}\ }\textbf {\bibinfo {volume} {4}}
  (\bibinfo {year} {2013})}\BibitemShut {NoStop}%
\bibitem [{\citenamefont {Ulm}\ \emph {et~al.}(2013)\citenamefont {Ulm},
  \citenamefont {Ro{\ss}nagel}, \citenamefont {Jacob}, \citenamefont
  {Degünther}, \citenamefont {Dawkins}, \citenamefont {Poschinger},
  \citenamefont {Nigmatullin}, \citenamefont {Retzker}, \citenamefont {Plenio},
  \citenamefont {Schmidt-Kaler},\ and\ \citenamefont {Singer}}]{Ulm2013a}%
  \BibitemOpen
  \bibfield  {author} {\bibinfo {author} {\bibfnamefont {S.}~\bibnamefont
  {Ulm}}, \bibinfo {author} {\bibfnamefont {J.}~\bibnamefont {Ro{\ss}nagel}},
  \bibinfo {author} {\bibfnamefont {G.}~\bibnamefont {Jacob}}, \bibinfo
  {author} {\bibfnamefont {C.}~\bibnamefont {Degünther}}, \bibinfo {author}
  {\bibfnamefont {S.~T.}\ \bibnamefont {Dawkins}}, \bibinfo {author}
  {\bibfnamefont {U.~G.}\ \bibnamefont {Poschinger}}, \bibinfo {author}
  {\bibfnamefont {R.}~\bibnamefont {Nigmatullin}}, \bibinfo {author}
  {\bibfnamefont {A.}~\bibnamefont {Retzker}}, \bibinfo {author} {\bibfnamefont
  {M.~B.}\ \bibnamefont {Plenio}}, \bibinfo {author} {\bibfnamefont
  {F.}~\bibnamefont {Schmidt-Kaler}}, \ and\ \bibinfo {author} {\bibfnamefont
  {K.}~\bibnamefont {Singer}},\ }\href {http://dx.doi.org/10.1038/ncomms3290}
  {\bibfield  {journal} {\bibinfo  {journal} {Nat Commun}\ }\textbf {\bibinfo
  {volume} {4}} (\bibinfo {year} {2013})}\BibitemShut {NoStop}%
\bibitem [{\citenamefont {Plenio}\ and\ \citenamefont
  {Retzker}(2013)}]{Plenio2013a}%
  \BibitemOpen
  \bibfield  {author} {\bibinfo {author} {\bibfnamefont {M.~B.}\ \bibnamefont
  {Plenio}}\ and\ \bibinfo {author} {\bibfnamefont {A.}~\bibnamefont
  {Retzker}},\ }\href {\doibase 10.1002/andp.201300740} {\bibfield  {journal}
  {\bibinfo  {journal} {Annalen der Physik}\ }\textbf {\bibinfo {volume}
  {525}},\ \bibinfo {pages} {A159} (\bibinfo {year} {2013})}\BibitemShut
  {NoStop}%
\bibitem [{\citenamefont {Bylinskii}\ \emph {et~al.}(2015)\citenamefont
  {Bylinskii}, \citenamefont {Gangloff},\ and\ \citenamefont
  {Vuletic}}]{Bylinskii2015a}%
  \BibitemOpen
  \bibfield  {author} {\bibinfo {author} {\bibfnamefont {A.}~\bibnamefont
  {Bylinskii}}, \bibinfo {author} {\bibfnamefont {D.}~\bibnamefont {Gangloff}},
  \ and\ \bibinfo {author} {\bibfnamefont {V.}~\bibnamefont {Vuletic}},\ }\href
  {\doibase 10.1126/science.1261422} {\bibfield  {journal} {\bibinfo  {journal}
  {Science}\ }\textbf {\bibinfo {volume} {348}},\ \bibinfo {pages} {1115}
  (\bibinfo {year} {2015})}\BibitemShut {NoStop}%
\bibitem [{\citenamefont {Diedrich}\ \emph {et~al.}(1989)\citenamefont
  {Diedrich}, \citenamefont {Bergquist}, \citenamefont {Itano},\ and\
  \citenamefont {Wineland}}]{Diedrich1989a}%
  \BibitemOpen
  \bibfield  {author} {\bibinfo {author} {\bibfnamefont {F.}~\bibnamefont
  {Diedrich}}, \bibinfo {author} {\bibfnamefont {J.~C.}\ \bibnamefont
  {Bergquist}}, \bibinfo {author} {\bibfnamefont {W.~M.}\ \bibnamefont
  {Itano}}, \ and\ \bibinfo {author} {\bibfnamefont {D.~J.}\ \bibnamefont
  {Wineland}},\ }\href {\doibase 10.1103/PhysRevLett.62.403} {\bibfield
  {journal} {\bibinfo  {journal} {Phys. Rev. Lett.}\ }\textbf {\bibinfo
  {volume} {62}},\ \bibinfo {pages} {403} (\bibinfo {year} {1989})}\BibitemShut
  {NoStop}%
\bibitem [{\citenamefont {Monroe}\ \emph {et~al.}(1995)\citenamefont {Monroe},
  \citenamefont {Meekhof}, \citenamefont {King}, \citenamefont {Jefferts},
  \citenamefont {Itano}, \citenamefont {Wineland},\ and\ \citenamefont
  {Gould}}]{Monroe1995a}%
  \BibitemOpen
  \bibfield  {author} {\bibinfo {author} {\bibfnamefont {C.}~\bibnamefont
  {Monroe}}, \bibinfo {author} {\bibfnamefont {D.~M.}\ \bibnamefont {Meekhof}},
  \bibinfo {author} {\bibfnamefont {B.~E.}\ \bibnamefont {King}}, \bibinfo
  {author} {\bibfnamefont {S.~R.}\ \bibnamefont {Jefferts}}, \bibinfo {author}
  {\bibfnamefont {W.~M.}\ \bibnamefont {Itano}}, \bibinfo {author}
  {\bibfnamefont {D.~J.}\ \bibnamefont {Wineland}}, \ and\ \bibinfo {author}
  {\bibfnamefont {P.}~\bibnamefont {Gould}},\ }\href {\doibase
  10.1103/PhysRevLett.75.4011} {\bibfield  {journal} {\bibinfo  {journal}
  {Phys. Rev. Lett.}\ }\textbf {\bibinfo {volume} {75}},\ \bibinfo {pages}
  {4011} (\bibinfo {year} {1995})}\BibitemShut {NoStop}%
\bibitem [{\citenamefont {King}\ \emph {et~al.}(1998)\citenamefont {King},
  \citenamefont {Wood}, \citenamefont {Myatt}, \citenamefont {Turchette},
  \citenamefont {Leibfried}, \citenamefont {Itano}, \citenamefont {Monroe},\
  and\ \citenamefont {Wineland}}]{King1998a}%
  \BibitemOpen
  \bibfield  {author} {\bibinfo {author} {\bibfnamefont {B.~E.}\ \bibnamefont
  {King}}, \bibinfo {author} {\bibfnamefont {C.~S.}\ \bibnamefont {Wood}},
  \bibinfo {author} {\bibfnamefont {C.~J.}\ \bibnamefont {Myatt}}, \bibinfo
  {author} {\bibfnamefont {Q.~A.}\ \bibnamefont {Turchette}}, \bibinfo {author}
  {\bibfnamefont {D.}~\bibnamefont {Leibfried}}, \bibinfo {author}
  {\bibfnamefont {W.~M.}\ \bibnamefont {Itano}}, \bibinfo {author}
  {\bibfnamefont {C.}~\bibnamefont {Monroe}}, \ and\ \bibinfo {author}
  {\bibfnamefont {D.~J.}\ \bibnamefont {Wineland}},\ }\href {\doibase
  10.1103/PhysRevLett.81.1525} {\bibfield  {journal} {\bibinfo  {journal}
  {Phys. Rev. Lett.}\ }\textbf {\bibinfo {volume} {81}},\ \bibinfo {pages}
  {1525} (\bibinfo {year} {1998})}\BibitemShut {NoStop}%
\bibitem [{\citenamefont {Roos}\ \emph {et~al.}(2000)\citenamefont {Roos},
  \citenamefont {Leibfried}, \citenamefont {Mundt}, \citenamefont
  {Schmidt-Kaler}, \citenamefont {Eschner},\ and\ \citenamefont
  {Blatt}}]{Roos2000a}%
  \BibitemOpen
  \bibfield  {author} {\bibinfo {author} {\bibfnamefont {C.~F.}\ \bibnamefont
  {Roos}}, \bibinfo {author} {\bibfnamefont {D.}~\bibnamefont {Leibfried}},
  \bibinfo {author} {\bibfnamefont {A.}~\bibnamefont {Mundt}}, \bibinfo
  {author} {\bibfnamefont {F.}~\bibnamefont {Schmidt-Kaler}}, \bibinfo {author}
  {\bibfnamefont {J.}~\bibnamefont {Eschner}}, \ and\ \bibinfo {author}
  {\bibfnamefont {R.}~\bibnamefont {Blatt}},\ }\href {\doibase
  10.1103/PhysRevLett.85.5547} {\bibfield  {journal} {\bibinfo  {journal}
  {Phys. Rev. Lett.}\ }\textbf {\bibinfo {volume} {85}},\ \bibinfo {pages}
  {5547} (\bibinfo {year} {2000})}\BibitemShut {NoStop}%
\bibitem [{\citenamefont {Lin}\ \emph {et~al.}(2013)\citenamefont {Lin},
  \citenamefont {Gaebler}, \citenamefont {Tan}, \citenamefont {Bowler},
  \citenamefont {Jost}, \citenamefont {Leibfried},\ and\ \citenamefont
  {Wineland}}]{Lin2013}%
  \BibitemOpen
  \bibfield  {author} {\bibinfo {author} {\bibfnamefont {Y.}~\bibnamefont
  {Lin}}, \bibinfo {author} {\bibfnamefont {J.~P.}\ \bibnamefont {Gaebler}},
  \bibinfo {author} {\bibfnamefont {T.~R.}\ \bibnamefont {Tan}}, \bibinfo
  {author} {\bibfnamefont {R.}~\bibnamefont {Bowler}}, \bibinfo {author}
  {\bibfnamefont {J.~D.}\ \bibnamefont {Jost}}, \bibinfo {author}
  {\bibfnamefont {D.}~\bibnamefont {Leibfried}}, \ and\ \bibinfo {author}
  {\bibfnamefont {D.~J.}\ \bibnamefont {Wineland}},\ }\href {\doibase
  10.1103/PhysRevLett.110.153002} {\bibfield  {journal} {\bibinfo  {journal}
  {Phys. Rev. Lett.}\ }\textbf {\bibinfo {volume} {110}},\ \bibinfo {pages}
  {153002} (\bibinfo {year} {2013})}\BibitemShut {NoStop}%
\bibitem [{\citenamefont {Leibfried}\ \emph {et~al.}(2003)\citenamefont
  {Leibfried}, \citenamefont {Blatt}, \citenamefont {Monroe},\ and\
  \citenamefont {Wineland}}]{Leibfried2003a}%
  \BibitemOpen
  \bibfield  {author} {\bibinfo {author} {\bibfnamefont {D.}~\bibnamefont
  {Leibfried}}, \bibinfo {author} {\bibfnamefont {R.}~\bibnamefont {Blatt}},
  \bibinfo {author} {\bibfnamefont {C.}~\bibnamefont {Monroe}}, \ and\ \bibinfo
  {author} {\bibfnamefont {D.}~\bibnamefont {Wineland}},\ }\href {\doibase
  10.1103/RevModPhys.75.281} {\bibfield  {journal} {\bibinfo  {journal} {Rev.
  Mod. Phys.}\ }\textbf {\bibinfo {volume} {75}},\ \bibinfo {pages} {281}
  (\bibinfo {year} {2003})}\BibitemShut {NoStop}%
\bibitem [{\citenamefont {Eschner}\ \emph {et~al.}(2003)\citenamefont
  {Eschner}, \citenamefont {Morigi}, \citenamefont {Schmidt-Kaler},\ and\
  \citenamefont {Blatt}}]{Eschner2003a}%
  \BibitemOpen
  \bibfield  {author} {\bibinfo {author} {\bibfnamefont {J.}~\bibnamefont
  {Eschner}}, \bibinfo {author} {\bibfnamefont {G.}~\bibnamefont {Morigi}},
  \bibinfo {author} {\bibfnamefont {F.}~\bibnamefont {Schmidt-Kaler}}, \ and\
  \bibinfo {author} {\bibfnamefont {R.}~\bibnamefont {Blatt}},\ }\href
  {\doibase 10.1364/JOSAB.20.001003} {\bibfield  {journal} {\bibinfo  {journal}
  {J. Opt. Soc. Am. B}\ }\textbf {\bibinfo {volume} {20}},\ \bibinfo {pages}
  {1003} (\bibinfo {year} {2003})}\BibitemShut {NoStop}%
\bibitem [{\citenamefont {Dalibard}\ and\ \citenamefont
  {Cohen-Tannoudji}(1989)}]{Dalibard1989a}%
  \BibitemOpen
  \bibfield  {author} {\bibinfo {author} {\bibfnamefont {J.}~\bibnamefont
  {Dalibard}}\ and\ \bibinfo {author} {\bibfnamefont {C.}~\bibnamefont
  {Cohen-Tannoudji}},\ }\href {\doibase 10.1364/JOSAB.6.002023} {\bibfield
  {journal} {\bibinfo  {journal} {J. Opt. Soc. Am. B}\ }\textbf {\bibinfo
  {volume} {6}},\ \bibinfo {pages} {2023} (\bibinfo {year} {1989})}\BibitemShut
  {NoStop}%
\bibitem [{\citenamefont {Ungar}\ \emph {et~al.}(1989)\citenamefont {Ungar},
  \citenamefont {Weiss}, \citenamefont {Riis},\ and\ \citenamefont
  {Chu}}]{Ungar1989a}%
  \BibitemOpen
  \bibfield  {author} {\bibinfo {author} {\bibfnamefont {P.~J.}\ \bibnamefont
  {Ungar}}, \bibinfo {author} {\bibfnamefont {D.~S.}\ \bibnamefont {Weiss}},
  \bibinfo {author} {\bibfnamefont {E.}~\bibnamefont {Riis}}, \ and\ \bibinfo
  {author} {\bibfnamefont {S.}~\bibnamefont {Chu}},\ }\href {\doibase
  10.1364/JOSAB.6.002058} {\bibfield  {journal} {\bibinfo  {journal} {J. Opt.
  Soc. Am. B}\ }\textbf {\bibinfo {volume} {6}},\ \bibinfo {pages} {2058}
  (\bibinfo {year} {1989})}\BibitemShut {NoStop}%
\bibitem [{\citenamefont {Lett}\ \emph {et~al.}(1988)\citenamefont {Lett},
  \citenamefont {Watts}, \citenamefont {Westbrook}, \citenamefont {Phillips},
  \citenamefont {Gould},\ and\ \citenamefont {Metcalf}}]{Lett1988a}%
  \BibitemOpen
  \bibfield  {author} {\bibinfo {author} {\bibfnamefont {P.~D.}\ \bibnamefont
  {Lett}}, \bibinfo {author} {\bibfnamefont {R.~N.}\ \bibnamefont {Watts}},
  \bibinfo {author} {\bibfnamefont {C.~I.}\ \bibnamefont {Westbrook}}, \bibinfo
  {author} {\bibfnamefont {W.~D.}\ \bibnamefont {Phillips}}, \bibinfo {author}
  {\bibfnamefont {P.~L.}\ \bibnamefont {Gould}}, \ and\ \bibinfo {author}
  {\bibfnamefont {H.~J.}\ \bibnamefont {Metcalf}},\ }\href {\doibase
  10.1103/PhysRevLett.61.169} {\bibfield  {journal} {\bibinfo  {journal} {Phys.
  Rev. Lett.}\ }\textbf {\bibinfo {volume} {61}},\ \bibinfo {pages} {169}
  (\bibinfo {year} {1988})}\BibitemShut {NoStop}%
\bibitem [{\citenamefont {Metcalf}\ and\ \citenamefont {van~der
  Straten}(1999)}]{Metcalf}%
  \BibitemOpen
  \bibfield  {author} {\bibinfo {author} {\bibfnamefont {H.~J.}\ \bibnamefont
  {Metcalf}}\ and\ \bibinfo {author} {\bibfnamefont {P.}~\bibnamefont {van~der
  Straten}},\ }\href@noop {} {\emph {\bibinfo {title} {Laser cooling and
  trapping}}}\ (\bibinfo  {publisher} {Springer},\ \bibinfo {year}
  {1999})\BibitemShut {NoStop}%
\bibitem [{\citenamefont {Jessen}\ \emph {et~al.}(1992)\citenamefont {Jessen},
  \citenamefont {Gerz}, \citenamefont {Lett}, \citenamefont {Phillips},
  \citenamefont {Rolston}, \citenamefont {Spreeuw},\ and\ \citenamefont
  {Westbrook}}]{Jessen1992}%
  \BibitemOpen
  \bibfield  {author} {\bibinfo {author} {\bibfnamefont {P.~S.}\ \bibnamefont
  {Jessen}}, \bibinfo {author} {\bibfnamefont {C.}~\bibnamefont {Gerz}},
  \bibinfo {author} {\bibfnamefont {P.~D.}\ \bibnamefont {Lett}}, \bibinfo
  {author} {\bibfnamefont {W.~D.}\ \bibnamefont {Phillips}}, \bibinfo {author}
  {\bibfnamefont {S.~L.}\ \bibnamefont {Rolston}}, \bibinfo {author}
  {\bibfnamefont {R.~J.~C.}\ \bibnamefont {Spreeuw}}, \ and\ \bibinfo {author}
  {\bibfnamefont {C.~I.}\ \bibnamefont {Westbrook}},\ }\href {\doibase
  10.1103/PhysRevLett.69.49} {\bibfield  {journal} {\bibinfo  {journal} {Phys.
  Rev. Lett.}\ }\textbf {\bibinfo {volume} {69}},\ \bibinfo {pages} {49}
  (\bibinfo {year} {1992})}\BibitemShut {NoStop}%
\bibitem [{\citenamefont {Verkerk}\ \emph {et~al.}(1992)\citenamefont
  {Verkerk}, \citenamefont {Lounis}, \citenamefont {Salomon}, \citenamefont
  {Cohen-Tannoudji}, \citenamefont {Courtois},\ and\ \citenamefont
  {Grynberg}}]{Verkerk1992}%
  \BibitemOpen
  \bibfield  {author} {\bibinfo {author} {\bibfnamefont {P.}~\bibnamefont
  {Verkerk}}, \bibinfo {author} {\bibfnamefont {B.}~\bibnamefont {Lounis}},
  \bibinfo {author} {\bibfnamefont {C.}~\bibnamefont {Salomon}}, \bibinfo
  {author} {\bibfnamefont {C.}~\bibnamefont {Cohen-Tannoudji}}, \bibinfo
  {author} {\bibfnamefont {J.-Y.}\ \bibnamefont {Courtois}}, \ and\ \bibinfo
  {author} {\bibfnamefont {G.}~\bibnamefont {Grynberg}},\ }\href {\doibase
  10.1103/PhysRevLett.68.3861} {\bibfield  {journal} {\bibinfo  {journal}
  {Phys. Rev. Lett.}\ }\textbf {\bibinfo {volume} {68}},\ \bibinfo {pages}
  {3861} (\bibinfo {year} {1992})}\BibitemShut {NoStop}%
\bibitem [{\citenamefont {Kastberg}\ \emph {et~al.}(1995)\citenamefont
  {Kastberg}, \citenamefont {Phillips}, \citenamefont {Rolston}, \citenamefont
  {Spreeuw},\ and\ \citenamefont {Jessen}}]{Kastberg1995}%
  \BibitemOpen
  \bibfield  {author} {\bibinfo {author} {\bibfnamefont {A.}~\bibnamefont
  {Kastberg}}, \bibinfo {author} {\bibfnamefont {W.~D.}\ \bibnamefont
  {Phillips}}, \bibinfo {author} {\bibfnamefont {S.~L.}\ \bibnamefont
  {Rolston}}, \bibinfo {author} {\bibfnamefont {R.~J.~C.}\ \bibnamefont
  {Spreeuw}}, \ and\ \bibinfo {author} {\bibfnamefont {P.~S.}\ \bibnamefont
  {Jessen}},\ }\href {\doibase 10.1103/PhysRevLett.74.1542} {\bibfield
  {journal} {\bibinfo  {journal} {Phys. Rev. Lett.}\ }\textbf {\bibinfo
  {volume} {74}},\ \bibinfo {pages} {1542} (\bibinfo {year}
  {1995})}\BibitemShut {NoStop}%
\bibitem [{\citenamefont {Winoto}\ \emph {et~al.}(1999)\citenamefont {Winoto},
  \citenamefont {DePue}, \citenamefont {Bramall},\ and\ \citenamefont
  {Weiss}}]{Winoto1999}%
  \BibitemOpen
  \bibfield  {author} {\bibinfo {author} {\bibfnamefont {S.~L.}\ \bibnamefont
  {Winoto}}, \bibinfo {author} {\bibfnamefont {M.~T.}\ \bibnamefont {DePue}},
  \bibinfo {author} {\bibfnamefont {N.~E.}\ \bibnamefont {Bramall}}, \ and\
  \bibinfo {author} {\bibfnamefont {D.~S.}\ \bibnamefont {Weiss}},\ }\href
  {\doibase 10.1103/PhysRevA.59.R19} {\bibfield  {journal} {\bibinfo  {journal}
  {Phys. Rev. A}\ }\textbf {\bibinfo {volume} {59}},\ \bibinfo {pages} {R19}
  (\bibinfo {year} {1999})}\BibitemShut {NoStop}%
\bibitem [{\citenamefont {Bakr}\ \emph {et~al.}(2010)\citenamefont {Bakr},
  \citenamefont {Peng}, \citenamefont {Tai}, \citenamefont {Ma}, \citenamefont
  {Simon}, \citenamefont {Gillen}, \citenamefont {Fölling}, \citenamefont
  {Pollet},\ and\ \citenamefont {Greiner}}]{Bakr2010}%
  \BibitemOpen
  \bibfield  {author} {\bibinfo {author} {\bibfnamefont {W.~S.}\ \bibnamefont
  {Bakr}}, \bibinfo {author} {\bibfnamefont {A.}~\bibnamefont {Peng}}, \bibinfo
  {author} {\bibfnamefont {M.~E.}\ \bibnamefont {Tai}}, \bibinfo {author}
  {\bibfnamefont {R.}~\bibnamefont {Ma}}, \bibinfo {author} {\bibfnamefont
  {J.}~\bibnamefont {Simon}}, \bibinfo {author} {\bibfnamefont {J.~I.}\
  \bibnamefont {Gillen}}, \bibinfo {author} {\bibfnamefont {S.}~\bibnamefont
  {Fölling}}, \bibinfo {author} {\bibfnamefont {L.}~\bibnamefont {Pollet}}, \
  and\ \bibinfo {author} {\bibfnamefont {M.}~\bibnamefont {Greiner}},\ }\href
  {\doibase 10.1126/science.1192368} {\bibfield  {journal} {\bibinfo  {journal}
  {Science}\ }\textbf {\bibinfo {volume} {329}},\ \bibinfo {pages} {547}
  (\bibinfo {year} {2010})}\BibitemShut {NoStop}%
\bibitem [{\citenamefont {Nuszmann}\ \emph {et~al.}(2005)\citenamefont
  {Nuszmann}, \citenamefont {Murr}, \citenamefont {Hijlkema}, \citenamefont
  {Weber}, \citenamefont {Kuhn},\ and\ \citenamefont {Rempe}}]{Nuszmann2005a}%
  \BibitemOpen
  \bibfield  {author} {\bibinfo {author} {\bibfnamefont {S.}~\bibnamefont
  {Nuszmann}}, \bibinfo {author} {\bibfnamefont {K.}~\bibnamefont {Murr}},
  \bibinfo {author} {\bibfnamefont {M.}~\bibnamefont {Hijlkema}}, \bibinfo
  {author} {\bibfnamefont {B.}~\bibnamefont {Weber}}, \bibinfo {author}
  {\bibfnamefont {A.}~\bibnamefont {Kuhn}}, \ and\ \bibinfo {author}
  {\bibfnamefont {G.}~\bibnamefont {Rempe}},\ }\href {\doibase
  10.1038/nphys120} {\bibfield  {journal} {\bibinfo  {journal} {Nat Phys}\
  }\textbf {\bibinfo {volume} {1}},\ \bibinfo {pages} {122} (\bibinfo {year}
  {2005})}\BibitemShut {NoStop}%
\bibitem [{\citenamefont {Boozer}\ \emph {et~al.}(2006)\citenamefont {Boozer},
  \citenamefont {Boca}, \citenamefont {Miller}, \citenamefont {Northup},\ and\
  \citenamefont {Kimble}}]{Boozer2006}%
  \BibitemOpen
  \bibfield  {author} {\bibinfo {author} {\bibfnamefont {A.~D.}\ \bibnamefont
  {Boozer}}, \bibinfo {author} {\bibfnamefont {A.}~\bibnamefont {Boca}},
  \bibinfo {author} {\bibfnamefont {R.}~\bibnamefont {Miller}}, \bibinfo
  {author} {\bibfnamefont {T.~E.}\ \bibnamefont {Northup}}, \ and\ \bibinfo
  {author} {\bibfnamefont {H.~J.}\ \bibnamefont {Kimble}},\ }\href {\doibase
  10.1103/PhysRevLett.97.083602} {\bibfield  {journal} {\bibinfo  {journal}
  {Phys. Rev. Lett.}\ }\textbf {\bibinfo {volume} {97}},\ \bibinfo {pages}
  {083602} (\bibinfo {year} {2006})}\BibitemShut {NoStop}%
\bibitem [{\citenamefont {Schlosser}\ \emph {et~al.}(2001)\citenamefont
  {Schlosser}, \citenamefont {Reymond}, \citenamefont {Protsenko},\ and\
  \citenamefont {Grangier}}]{schlosser2001}%
  \BibitemOpen
  \bibfield  {author} {\bibinfo {author} {\bibfnamefont {N.}~\bibnamefont
  {Schlosser}}, \bibinfo {author} {\bibfnamefont {G.}~\bibnamefont {Reymond}},
  \bibinfo {author} {\bibfnamefont {I.}~\bibnamefont {Protsenko}}, \ and\
  \bibinfo {author} {\bibfnamefont {P.}~\bibnamefont {Grangier}},\ }\href
  {\doibase 10.1038/35082512} {\bibfield  {journal} {\bibinfo  {journal}
  {Nature}\ }\textbf {\bibinfo {volume} {411}},\ \bibinfo {pages} {1024}
  (\bibinfo {year} {2001})}\BibitemShut {NoStop}%
\bibitem [{\citenamefont {Kaufman}\ \emph {et~al.}(2012)\citenamefont
  {Kaufman}, \citenamefont {Lester},\ and\ \citenamefont
  {Regal}}]{Kaufman2012a}%
  \BibitemOpen
  \bibfield  {author} {\bibinfo {author} {\bibfnamefont {A.~M.}\ \bibnamefont
  {Kaufman}}, \bibinfo {author} {\bibfnamefont {B.~J.}\ \bibnamefont {Lester}},
  \ and\ \bibinfo {author} {\bibfnamefont {C.~A.}\ \bibnamefont {Regal}},\
  }\href {\doibase 10.1103/PhysRevX.2.041014} {\bibfield  {journal} {\bibinfo
  {journal} {Phys. Rev. X}\ }\textbf {\bibinfo {volume} {2}},\ \bibinfo {pages}
  {041014} (\bibinfo {year} {2012})}\BibitemShut {NoStop}%
\bibitem [{\citenamefont {McGovern}\ \emph {et~al.}(2011)\citenamefont
  {McGovern}, \citenamefont {Hilliard}, \citenamefont {Gr\"{u}nzweig},\ and\
  \citenamefont {Andersen}}]{McGovern2011}%
  \BibitemOpen
  \bibfield  {author} {\bibinfo {author} {\bibfnamefont {M.}~\bibnamefont
  {McGovern}}, \bibinfo {author} {\bibfnamefont {A.~J.}\ \bibnamefont
  {Hilliard}}, \bibinfo {author} {\bibfnamefont {T.}~\bibnamefont
  {Gr\"{u}nzweig}}, \ and\ \bibinfo {author} {\bibfnamefont {M.~F.}\
  \bibnamefont {Andersen}},\ }\href {\doibase 10.1364/OL.36.001041} {\bibfield
  {journal} {\bibinfo  {journal} {Opt. Lett.}\ }\textbf {\bibinfo {volume}
  {36}},\ \bibinfo {pages} {1041} (\bibinfo {year} {2011})}\BibitemShut
  {NoStop}%
\bibitem [{\citenamefont {Wineland}\ \emph {et~al.}(1992)\citenamefont
  {Wineland}, \citenamefont {Dalibard},\ and\ \citenamefont
  {Cohen-Tannoudji}}]{Wineland1992a}%
  \BibitemOpen
  \bibfield  {author} {\bibinfo {author} {\bibfnamefont {D.~J.}\ \bibnamefont
  {Wineland}}, \bibinfo {author} {\bibfnamefont {J.}~\bibnamefont {Dalibard}},
  \ and\ \bibinfo {author} {\bibfnamefont {C.}~\bibnamefont
  {Cohen-Tannoudji}},\ }\href {\doibase 10.1364/JOSAB.9.000032} {\bibfield
  {journal} {\bibinfo  {journal} {J. Opt. Soc. Am. B}\ }\textbf {\bibinfo
  {volume} {9}},\ \bibinfo {pages} {32} (\bibinfo {year} {1992})}\BibitemShut
  {NoStop}%
\bibitem [{\citenamefont {Cirac}\ \emph {et~al.}(1992)\citenamefont {Cirac},
  \citenamefont {Blatt}, \citenamefont {Zoller},\ and\ \citenamefont
  {Phillips}}]{Cirac1992a}%
  \BibitemOpen
  \bibfield  {author} {\bibinfo {author} {\bibfnamefont {J.~I.}\ \bibnamefont
  {Cirac}}, \bibinfo {author} {\bibfnamefont {R.}~\bibnamefont {Blatt}},
  \bibinfo {author} {\bibfnamefont {P.}~\bibnamefont {Zoller}}, \ and\ \bibinfo
  {author} {\bibfnamefont {W.~D.}\ \bibnamefont {Phillips}},\ }\href {\doibase
  10.1103/PhysRevA.46.2668} {\bibfield  {journal} {\bibinfo  {journal} {Phys.
  Rev. A}\ }\textbf {\bibinfo {volume} {46}},\ \bibinfo {pages} {2668}
  (\bibinfo {year} {1992})}\BibitemShut {NoStop}%
\bibitem [{\citenamefont {Cirac}\ \emph {et~al.}(1993)\citenamefont {Cirac},
  \citenamefont {Blatt}, \citenamefont {Parkins},\ and\ \citenamefont
  {Zoller}}]{Cirac1993a}%
  \BibitemOpen
  \bibfield  {author} {\bibinfo {author} {\bibfnamefont {J.~I.}\ \bibnamefont
  {Cirac}}, \bibinfo {author} {\bibfnamefont {R.}~\bibnamefont {Blatt}},
  \bibinfo {author} {\bibfnamefont {A.~S.}\ \bibnamefont {Parkins}}, \ and\
  \bibinfo {author} {\bibfnamefont {P.}~\bibnamefont {Zoller}},\ }\href
  {\doibase 10.1103/PhysRevA.48.1434} {\bibfield  {journal} {\bibinfo
  {journal} {Phys. Rev. A}\ }\textbf {\bibinfo {volume} {48}},\ \bibinfo
  {pages} {1434} (\bibinfo {year} {1993})}\BibitemShut {NoStop}%
\bibitem [{\citenamefont {Yoo}\ and\ \citenamefont
  {Javanainen}(1993)}]{Yoo1993a}%
  \BibitemOpen
  \bibfield  {author} {\bibinfo {author} {\bibfnamefont {S.~M.}\ \bibnamefont
  {Yoo}}\ and\ \bibinfo {author} {\bibfnamefont {J.}~\bibnamefont
  {Javanainen}},\ }\href {\doibase 10.1103/PhysRevA.48.R30} {\bibfield
  {journal} {\bibinfo  {journal} {Phys. Rev. A}\ }\textbf {\bibinfo {volume}
  {48}},\ \bibinfo {pages} {R30} (\bibinfo {year} {1993})}\BibitemShut
  {NoStop}%
\bibitem [{\citenamefont {Li}\ and\ \citenamefont {M{\o}lmer}(1994)}]{Li1993a}%
  \BibitemOpen
  \bibfield  {author} {\bibinfo {author} {\bibfnamefont {Y.}~\bibnamefont
  {Li}}\ and\ \bibinfo {author} {\bibnamefont {M{\o}lmer}},\ }\href@noop {}
  {\bibfield  {journal} {\bibinfo  {journal} {Laser Physics}\ }\textbf
  {\bibinfo {volume} {4}},\ \bibinfo {pages} {829} (\bibinfo {year}
  {1994})}\BibitemShut {NoStop}%
\bibitem [{\citenamefont {Birkl}\ \emph {et~al.}(1994)\citenamefont {Birkl},
  \citenamefont {Yeazell}, \citenamefont {R\"{u}ckerl},\ and\ \citenamefont
  {Walther}}]{Birkl1994a}%
  \BibitemOpen
  \bibfield  {author} {\bibinfo {author} {\bibfnamefont {G.}~\bibnamefont
  {Birkl}}, \bibinfo {author} {\bibfnamefont {J.~A.}\ \bibnamefont {Yeazell}},
  \bibinfo {author} {\bibfnamefont {R.}~\bibnamefont {R\"{u}ckerl}}, \ and\
  \bibinfo {author} {\bibfnamefont {H.}~\bibnamefont {Walther}},\ }\href
  {http://stacks.iop.org/0295-5075/27/i=3/a=005} {\bibfield  {journal}
  {\bibinfo  {journal} {Europhys. Lett.}\ }\textbf {\bibinfo {volume} {27}},\
  \bibinfo {pages} {197} (\bibinfo {year} {1994})}\BibitemShut {NoStop}%
\bibitem [{\citenamefont {van~der Straten}\ \emph {et~al.}(1993)\citenamefont
  {van~der Straten}, \citenamefont {Shang}, \citenamefont {Sheehy},
  \citenamefont {Metcalf},\ and\ \citenamefont {Nienhuis}}]{Straten1993a}%
  \BibitemOpen
  \bibfield  {author} {\bibinfo {author} {\bibfnamefont {P.}~\bibnamefont
  {van~der Straten}}, \bibinfo {author} {\bibfnamefont {S.-Q.}\ \bibnamefont
  {Shang}}, \bibinfo {author} {\bibfnamefont {B.}~\bibnamefont {Sheehy}},
  \bibinfo {author} {\bibfnamefont {H.}~\bibnamefont {Metcalf}}, \ and\
  \bibinfo {author} {\bibfnamefont {G.}~\bibnamefont {Nienhuis}},\ }\href
  {\doibase 10.1103/PhysRevA.47.4160} {\bibfield  {journal} {\bibinfo
  {journal} {Phys. Rev. A}\ }\textbf {\bibinfo {volume} {47}},\ \bibinfo
  {pages} {4160} (\bibinfo {year} {1993})}\BibitemShut {NoStop}%
\bibitem [{\citenamefont {Berkeland}\ and\ \citenamefont
  {Boshier}(2002)}]{Berkeland2002a}%
  \BibitemOpen
  \bibfield  {author} {\bibinfo {author} {\bibfnamefont {D.~J.}\ \bibnamefont
  {Berkeland}}\ and\ \bibinfo {author} {\bibfnamefont {M.~G.}\ \bibnamefont
  {Boshier}},\ }\href {\doibase 10.1103/PhysRevA.65.033413} {\bibfield
  {journal} {\bibinfo  {journal} {Phys. Rev. A}\ }\textbf {\bibinfo {volume}
  {65}},\ \bibinfo {pages} {033413} (\bibinfo {year} {2002})}\BibitemShut
  {NoStop}%
\bibitem [{\citenamefont {Ejtemaee}\ \emph {et~al.}(2010)\citenamefont
  {Ejtemaee}, \citenamefont {Thomas},\ and\ \citenamefont
  {Haljan}}]{Ejtemaee2010a}%
  \BibitemOpen
  \bibfield  {author} {\bibinfo {author} {\bibfnamefont {S.}~\bibnamefont
  {Ejtemaee}}, \bibinfo {author} {\bibfnamefont {R.}~\bibnamefont {Thomas}}, \
  and\ \bibinfo {author} {\bibfnamefont {P.~C.}\ \bibnamefont {Haljan}},\
  }\href {\doibase 10.1103/PhysRevA.82.063419} {\bibfield  {journal} {\bibinfo
  {journal} {Phys. Rev. A}\ }\textbf {\bibinfo {volume} {82}},\ \bibinfo
  {pages} {063419} (\bibinfo {year} {2010})}\BibitemShut {NoStop}%
\bibitem [{\citenamefont {Ejtemaee}\ and\ \citenamefont
  {Haljan}()}]{Ejtemaee2016b}%
  \BibitemOpen
  \bibfield  {author} {\bibinfo {author} {\bibfnamefont {S.}~\bibnamefont
  {Ejtemaee}}\ and\ \bibinfo {author} {\bibfnamefont {P.~C.}\ \bibnamefont
  {Haljan}},\ }\href@noop {} {\bibinfo  {journal} {in preparation}\
  }\BibitemShut {NoStop}%
\bibitem [{\citenamefont {Gordon}\ and\ \citenamefont
  {Ashkin}(1980)}]{Gordon1980a}%
  \BibitemOpen
\bibfield  {journal} {  }\bibfield  {author} {\bibinfo {author} {\bibfnamefont
  {J.~P.}\ \bibnamefont {Gordon}}\ and\ \bibinfo {author} {\bibfnamefont
  {A.}~\bibnamefont {Ashkin}},\ }\href {\doibase 10.1103/PhysRevA.21.1606}
  {\bibfield  {journal} {\bibinfo  {journal} {Phys. Rev. A}\ }\textbf {\bibinfo
  {volume} {21}},\ \bibinfo {pages} {1606} (\bibinfo {year}
  {1980})}\BibitemShut {NoStop}%
\bibitem [{\citenamefont {Nienhuis}\ \emph {et~al.}(1991)\citenamefont
  {Nienhuis}, \citenamefont {van~der Straten},\ and\ \citenamefont
  {Shang}}]{Nienhuis1991a}%
  \BibitemOpen
  \bibfield  {author} {\bibinfo {author} {\bibfnamefont {G.}~\bibnamefont
  {Nienhuis}}, \bibinfo {author} {\bibfnamefont {P.}~\bibnamefont {van~der
  Straten}}, \ and\ \bibinfo {author} {\bibfnamefont {S.-Q.}\ \bibnamefont
  {Shang}},\ }\href {\doibase 10.1103/PhysRevA.44.462} {\bibfield  {journal}
  {\bibinfo  {journal} {Phys. Rev. A}\ }\textbf {\bibinfo {volume} {44}},\
  \bibinfo {pages} {462} (\bibinfo {year} {1991})}\BibitemShut {NoStop}%
\bibitem [{\citenamefont {Dalibard}\ \emph {et~al.}(1992)\citenamefont
  {Dalibard}, \citenamefont {Castin},\ and\ \citenamefont
  {M\o{}lmer}}]{Dalibard1992a}%
  \BibitemOpen
  \bibfield  {author} {\bibinfo {author} {\bibfnamefont {J.}~\bibnamefont
  {Dalibard}}, \bibinfo {author} {\bibfnamefont {Y.}~\bibnamefont {Castin}}, \
  and\ \bibinfo {author} {\bibfnamefont {K.}~\bibnamefont {M\o{}lmer}},\ }\href
  {\doibase 10.1103/PhysRevLett.68.580} {\bibfield  {journal} {\bibinfo
  {journal} {Phys. Rev. Lett.}\ }\textbf {\bibinfo {volume} {68}},\ \bibinfo
  {pages} {580} (\bibinfo {year} {1992})}\BibitemShut {NoStop}%
\bibitem [{\citenamefont {Johansson}\ \emph {et~al.}(2013)\citenamefont
  {Johansson}, \citenamefont {Nation},\ and\ \citenamefont
  {Nori}}]{Johansson20131234}%
  \BibitemOpen
  \bibfield  {author} {\bibinfo {author} {\bibfnamefont {J.}~\bibnamefont
  {Johansson}}, \bibinfo {author} {\bibfnamefont {P.}~\bibnamefont {Nation}}, \
  and\ \bibinfo {author} {\bibfnamefont {F.}~\bibnamefont {Nori}},\ }\href
  {\doibase 10.1016/j.cpc.2012.11.019} {\bibfield  {journal} {\bibinfo
  {journal} {Computer Physics Communications}\ }\textbf {\bibinfo {volume}
  {184}},\ \bibinfo {pages} {1234 } (\bibinfo {year} {2013})}\BibitemShut
  {NoStop}%
\bibitem [{sid()}]{sidebandthermometry}%
  \BibitemOpen
  \href@noop {} {}\bibinfo {note} {This is equivalent to thermometry based on
  red and blue sidebands described in~\cite{Leibfried2003a}}\BibitemShut
  {NoStop}%
\bibitem [{\citenamefont {Retzker}\ \emph {et~al.}(2008)\citenamefont
  {Retzker}, \citenamefont {Thompson}, \citenamefont {Segal},\ and\
  \citenamefont {Plenio}}]{Retzker2008a}%
  \BibitemOpen
  \bibfield  {author} {\bibinfo {author} {\bibfnamefont {A.}~\bibnamefont
  {Retzker}}, \bibinfo {author} {\bibfnamefont {R.~C.}\ \bibnamefont
  {Thompson}}, \bibinfo {author} {\bibfnamefont {D.~M.}\ \bibnamefont {Segal}},
  \ and\ \bibinfo {author} {\bibfnamefont {M.~B.}\ \bibnamefont {Plenio}},\
  }\href {\doibase 10.1103/PhysRevLett.101.260504} {\bibfield  {journal}
  {\bibinfo  {journal} {Phys. Rev. Lett.}\ }\textbf {\bibinfo {volume} {101}},\
  \bibinfo {pages} {260504} (\bibinfo {year} {2008})}\BibitemShut {NoStop}%
\bibitem [{\citenamefont {Lin}\ and\ \citenamefont {Duan}(2011)}]{Lin2011}%
  \BibitemOpen
  \bibfield  {author} {\bibinfo {author} {\bibfnamefont {G.-D.}\ \bibnamefont
  {Lin}}\ and\ \bibinfo {author} {\bibfnamefont {L.-M.}\ \bibnamefont {Duan}},\
  }\href {\doibase 10.1088/1367-2630/13/7/075015} {\bibfield  {journal}
  {\bibinfo  {journal} {New Journal of Physics}\ }\textbf {\bibinfo {volume}
  {13}},\ \bibinfo {pages} {075015} (\bibinfo {year} {2011})}\BibitemShut
  {NoStop}%
\bibitem [{\citenamefont {Pruttivarasin}\ \emph {et~al.}(2011)\citenamefont
  {Pruttivarasin}, \citenamefont {Ramm}, \citenamefont {Talukdar},
  \citenamefont {Kreuter},\ and\ \citenamefont {Häffner}}]{Pruttivarasin2011}%
  \BibitemOpen
  \bibfield  {author} {\bibinfo {author} {\bibfnamefont {T.}~\bibnamefont
  {Pruttivarasin}}, \bibinfo {author} {\bibfnamefont {M.}~\bibnamefont {Ramm}},
  \bibinfo {author} {\bibfnamefont {I.}~\bibnamefont {Talukdar}}, \bibinfo
  {author} {\bibfnamefont {A.}~\bibnamefont {Kreuter}}, \ and\ \bibinfo
  {author} {\bibfnamefont {H.}~\bibnamefont {Häffner}},\ }\href {\doibase
  10.1088/1367-2630/13/7/075012} {\bibfield  {journal} {\bibinfo  {journal}
  {New Journal of Physics}\ }\textbf {\bibinfo {volume} {13}},\ \bibinfo
  {pages} {075012} (\bibinfo {year} {2011})}\BibitemShut {NoStop}%
\bibitem [{\citenamefont {Bermudez}\ \emph {et~al.}(2013)\citenamefont
  {Bermudez}, \citenamefont {Bruderer},\ and\ \citenamefont
  {Plenio}}]{Bermudez2013}%
  \BibitemOpen
  \bibfield  {author} {\bibinfo {author} {\bibfnamefont {A.}~\bibnamefont
  {Bermudez}}, \bibinfo {author} {\bibfnamefont {M.}~\bibnamefont {Bruderer}},
  \ and\ \bibinfo {author} {\bibfnamefont {M.~B.}\ \bibnamefont {Plenio}},\
  }\href {\doibase 10.1103/PhysRevLett.111.040601} {\bibfield  {journal}
  {\bibinfo  {journal} {Physical Review Letters}\ }\textbf {\bibinfo {volume}
  {111}},\ \bibinfo {pages} {040601} (\bibinfo {year} {2013})}\BibitemShut
  {NoStop}%
\bibitem [{\citenamefont {Freitas}\ \emph {et~al.}(2013)\citenamefont
  {Freitas}, \citenamefont {Martinez},\ and\ \citenamefont
  {Paz}}]{Freitas2013}%
  \BibitemOpen
  \bibfield  {author} {\bibinfo {author} {\bibfnamefont {N.}~\bibnamefont
  {Freitas}}, \bibinfo {author} {\bibfnamefont {E.}~\bibnamefont {Martinez}}, \
  and\ \bibinfo {author} {\bibfnamefont {J.~P.}\ \bibnamefont {Paz}},\ }\href
  {http://arxiv.org/abs/1312.6644} {\bibfield  {journal} {\bibinfo  {journal}
  {arXiv preprint arXiv:1312.6644}\ } (\bibinfo {year} {2013})}\BibitemShut
  {NoStop}%
\bibitem [{\citenamefont {Ruiz}\ \emph {et~al.}(2014)\citenamefont {Ruiz},
  \citenamefont {Alonso}, \citenamefont {Plenio},\ and\ \citenamefont {del
  Campo}}]{Ruiz2014}%
  \BibitemOpen
  \bibfield  {author} {\bibinfo {author} {\bibfnamefont {A.}~\bibnamefont
  {Ruiz}}, \bibinfo {author} {\bibfnamefont {D.}~\bibnamefont {Alonso}},
  \bibinfo {author} {\bibfnamefont {M.~B.}\ \bibnamefont {Plenio}}, \ and\
  \bibinfo {author} {\bibfnamefont {A.}~\bibnamefont {del Campo}},\ }\href
  {\doibase 10.1103/PhysRevB.89.214305} {\bibfield  {journal} {\bibinfo
  {journal} {Physical Review B}\ }\textbf {\bibinfo {volume} {89}},\ \bibinfo
  {pages} {214305} (\bibinfo {year} {2014})}\BibitemShut {NoStop}%
\bibitem [{\citenamefont {Ospelkaus}\ \emph {et~al.}(2011)\citenamefont
  {Ospelkaus}, \citenamefont {Warring}, \citenamefont {Colombe}, \citenamefont
  {Brown}, \citenamefont {Amini}, \citenamefont {Leibfried},\ and\
  \citenamefont {Wineland}}]{Ospelkaus2011}%
  \BibitemOpen
  \bibfield  {author} {\bibinfo {author} {\bibfnamefont {C.}~\bibnamefont
  {Ospelkaus}}, \bibinfo {author} {\bibfnamefont {U.}~\bibnamefont {Warring}},
  \bibinfo {author} {\bibfnamefont {Y.}~\bibnamefont {Colombe}}, \bibinfo
  {author} {\bibfnamefont {K.~R.}\ \bibnamefont {Brown}}, \bibinfo {author}
  {\bibfnamefont {J.~M.}\ \bibnamefont {Amini}}, \bibinfo {author}
  {\bibfnamefont {D.}~\bibnamefont {Leibfried}}, \ and\ \bibinfo {author}
  {\bibfnamefont {D.~J.}\ \bibnamefont {Wineland}},\ }\href
  {http://dx.doi.org/10.1038/nature10290} {\bibfield  {journal} {\bibinfo
  {journal} {Nature}\ }\textbf {\bibinfo {volume} {476}},\ \bibinfo {pages}
  {181} (\bibinfo {year} {2011})}\BibitemShut {NoStop}%
\bibitem [{\citenamefont {Timoney}\ \emph {et~al.}(2011)\citenamefont
  {Timoney}, \citenamefont {Baumgart}, \citenamefont {Johanning}, \citenamefont
  {Varon}, \citenamefont {Plenio}, \citenamefont {Retzker},\ and\ \citenamefont
  {Wunderlich}}]{Timoney2011}%
  \BibitemOpen
  \bibfield  {author} {\bibinfo {author} {\bibfnamefont {N.}~\bibnamefont
  {Timoney}}, \bibinfo {author} {\bibfnamefont {I.}~\bibnamefont {Baumgart}},
  \bibinfo {author} {\bibfnamefont {M.}~\bibnamefont {Johanning}}, \bibinfo
  {author} {\bibfnamefont {A.~F.}\ \bibnamefont {Varon}}, \bibinfo {author}
  {\bibfnamefont {M.~B.}\ \bibnamefont {Plenio}}, \bibinfo {author}
  {\bibfnamefont {A.}~\bibnamefont {Retzker}}, \ and\ \bibinfo {author}
  {\bibfnamefont {C.}~\bibnamefont {Wunderlich}},\ }\href
  {http://dx.doi.org/10.1038/nature10319} {\bibfield  {journal} {\bibinfo
  {journal} {Nature}\ }\textbf {\bibinfo {volume} {476}},\ \bibinfo {pages}
  {185} (\bibinfo {year} {2011})}\BibitemShut {NoStop}%
\bibitem [{\citenamefont {Webster}\ \emph {et~al.}(2013)\citenamefont
  {Webster}, \citenamefont {Weidt}, \citenamefont {Lake}, \citenamefont
  {McLoughlin},\ and\ \citenamefont {Hensinger}}]{Webster2013a}%
  \BibitemOpen
  \bibfield  {author} {\bibinfo {author} {\bibfnamefont {S.~C.}\ \bibnamefont
  {Webster}}, \bibinfo {author} {\bibfnamefont {S.}~\bibnamefont {Weidt}},
  \bibinfo {author} {\bibfnamefont {K.}~\bibnamefont {Lake}}, \bibinfo {author}
  {\bibfnamefont {J.~J.}\ \bibnamefont {McLoughlin}}, \ and\ \bibinfo {author}
  {\bibfnamefont {W.~K.}\ \bibnamefont {Hensinger}},\ }\href {\doibase
  10.1103/PhysRevLett.111.140501} {\bibfield  {journal} {\bibinfo  {journal}
  {Phys. Rev. Lett.}\ }\textbf {\bibinfo {volume} {111}},\ \bibinfo {pages}
  {140501} (\bibinfo {year} {2013})}\BibitemShut {NoStop}%
\bibitem [{\citenamefont {Lin}\ \emph {et~al.}(2009)\citenamefont {Lin},
  \citenamefont {Zhu}, \citenamefont {Islam}, \citenamefont {Kim},
  \citenamefont {Chang}, \citenamefont {Korenblit}, \citenamefont {Monroe},\
  and\ \citenamefont {Duan}}]{Lin2009a}%
  \BibitemOpen
  \bibfield  {author} {\bibinfo {author} {\bibfnamefont {G.-D.}\ \bibnamefont
  {Lin}}, \bibinfo {author} {\bibfnamefont {S.-L.}\ \bibnamefont {Zhu}},
  \bibinfo {author} {\bibfnamefont {R.}~\bibnamefont {Islam}}, \bibinfo
  {author} {\bibfnamefont {K.}~\bibnamefont {Kim}}, \bibinfo {author}
  {\bibfnamefont {M.-S.}\ \bibnamefont {Chang}}, \bibinfo {author}
  {\bibfnamefont {S.}~\bibnamefont {Korenblit}}, \bibinfo {author}
  {\bibfnamefont {C.}~\bibnamefont {Monroe}}, \ and\ \bibinfo {author}
  {\bibfnamefont {L.-M.}\ \bibnamefont {Duan}},\ }\href {\doibase
  10.1209/0295-5075/86/60004} {\bibfield  {journal} {\bibinfo  {journal}
  {Europhys. Lett.}\ }\textbf {\bibinfo {volume} {86}},\ \bibinfo {pages}
  {60004} (\bibinfo {year} {2009})}\BibitemShut {NoStop}%
\end{thebibliography}


%***************************figures*****************************

%%%%%%%%%%%%%%%%%%%%%%%%%%%%%%%%%%%%%%%%%%%%%%%%%%%%%%%%%%%%%%%%%%%%%%%%%%

\end{document}